\documentclass{aa}  

\usepackage[pdftex, colorlinks=true, linkcolor=blue, citecolor=blue, urlcolor=blue]{hyperref}
\usepackage{nameref,ulem}

\usepackage[dvipsnames]{xcolor}
\usepackage{threeparttable}

\newcommand{\mic}{$\mu$m}

\newcommand{\hk}{HK\,Tau\,B}
\newcommand{\hv}{HV\,Tau\,C}
\newcommand{\tauZero}{Tau\,042021}

\newcommand{\Haro}{Haro\,6-5B}

\newcommand{\EsoSixNine}{ESO\,H$\alpha$\,569}
\newcommand{\ESOSevFour}{ESO\,H$\alpha$\,574}
\newcommand{\hhq}{HH\,48\,NE}

\newcommand\finalfinal[1]{#1}
\newcommand\final[1]{#1}
\newcommand\rev[1]{#1}

\usepackage{graphicx}
\usepackage{txfonts}
\begin{document}

   \title{Observations of highly inclined disks with ALMA}

   \subtitle{Results from  $^{12}$CO gas and continuum observations}

   \author{L. Martinien \inst{1}
          \and
           G. Duch\^ene\inst{1,2}
          \and
          \'A. Ribas\inst{3,4}
          \and
          M. Villenave\inst{1}
          \and
          F. M\'enard\inst{1}
          \and 
          K.R. Stapelfeldt\inst{5,}\thanks{On sabbatical leave at IPAG}
           \and 
           C. Pinte\inst{1}}

   \institute{Univ. Grenoble Alpes, CNRS, IPAG, 38000 Grenoble, France \\
    \email{laurine.martinien@univ-grenoble-alpes.fr}
    \and Astronomy Department, University of California Berkeley, Berkeley CA 94720-3411, USA
    \and Astronomy Unit, Department of Physics and Astronomy, Queen Mary University of London, Mile End Road, London E1 4NS, UK
    \and Institute of Astronomy, University of Cambridge, Madingley Road, Cambridge, CB3 0HA, UK
    \and Jet Propulsion Laboratory, California Institute of Technology, 4800 Oak Grove Drive, Pasadena, CA 91109, USA}
   \date{Received ; accepted }

 
  \abstract
   {}
   {We aim to study the radial and vertical extents of $^{12}$CO gas, millimeter dust thermal emission and optical/near infrared scattered light by dust in highly inclined protoplanetary disks. These parameters are indicators of radial drift and vertical settling, essential for better constraining planet formation. Additionally, we aim to provide estimates of the dynamical stellar masses including those for which no prior measurements exist.}
   {We analyze a sample of 14 highly inclined protoplanetary disks for which the vertical extent of the emission layers can be constrained directly. We present ALMA high angular resolution band 7 (0.9\,mm) continuum images and $^{12}$CO (3-2) gas moment maps as well as HST and VLT/SPHERE scattered light images. We estimate the dynamical masses using position-velocity diagrams.}
   {The majority of disks in our sample (11 out of 14) follow $R_\mathrm{gas}$ > $R_\mathrm{dust,\,\mu m}$ > $R_\mathrm{dust, \, mm}$. The other 3 disks, including 2 in multiple systems, appear more extended in millimeter continuum than in scattered light. Highly inclined disks tend to appear less radially extended in CO gas line emission than in millimeter dust continuum compared to less inclined disks. This results from optical depth effects and/or radial drift. The known correlation between disk size and millimeter  \rev{continuum and line fluxes are confirmed in our sample with highly inclined disks significantly fainter than disks seen at lower inclination for a given disk radius. We found that this correlation is} significantly tightened once fluxes are corrected for the disk inclination, consistent with the disks being optically thick at millimeter wavelengths. Regarding the vertical extent defined as the apparent emitting height, most disks in our sample follow $H_\mathrm{gas}$ >  $H_\mathrm{dust , \, mm}$. This strengthens our previous findings that the millimeter dust is highly decoupled from the gas and forms a layer in the disk midplane due to vertical settling. Most disks appear more vertically extended in gas than in scattered light, suggesting that the $\mu$m-sized dust is not fully coupled to the gas. 
   We also estimated dynamical masses for the first time for most of the objects in our sample. We found an anti-correlation between the dynamical mass and the aspect ratio, emphasizing the dominant role of gravity in setting the disk vertical extent, but no correlation with the disk radius. 
   }
   {
   }


   \maketitle
%

\section{Introduction}

Planets form in gas-rich protoplanetary disks \citep{benisty2022optical, Pinte2023}.  The process leading to km-sized bodies and larger is a long-standing puzzle. The classic core-accretion scenario \citep{Pollack1996} that requires small $\mu$m-sized particles to grow up by collisions into large pebbles, boulders, and eventually planetesimals, remains the main framework for planet building today. However, given the short lifetimes of disks (typical 5 Myr, e.g. \citealp{Hernandez_2008}) some models requiring an increase of dust concentration in the midplane have been proposed to speed-up the grain growth up to planetesimal scales, e.g., streaming instability \citep{Youdin2005, Johansen2007} and pebble accretion \citep{Lambrechts2012}. 
Large mm-sized grains that experience gas drag and
are subject to vertical settling toward the midplane and inward radial drift \citep{Weidenschilling_1977, Barriere_Fouchet_2005}. Radial drift is predicted to be $\approx$ 100 times slower than vertical settling \citep{Laibe_2014}. 
Vertical settling results from the damping of gravity-induced vertical oscillations of dust particles, while radial drift arises from the azimuthal velocity difference between Keplerian dust and sub-Keplerian gas. 
However, the expected timescale for $\mu$m-sized grains, probed by scattered light, is longer than the disk lifetime time, implying that small grains should remain colocated with the gas. Protoplanetary disks appear larger radially in their gas component than in their mm-sized dust component \citep[e.g.][]{Ansdell_2018}. This can be explained by radial drift. However, these studies were biased toward low- or mid-inclination disks for which low optical depth in the outer disk \citep{Andrews2020} and limited sensitivity can affect detectability, and therefore interpretation. As a result, these processes are poorly constrained. 

In order to constrain vertical settling and radial drift, it is important to quantify the radial and vertical distribution of dust of different sizes and compare it to the gas extent. Indeed, the dust extent is regulated by grain growth and the degree of migration in the disk, while the gas extent is dominated by physical and chemical interactions and dynamics of the disk. This gas extent is preferentially measured with the $^{12}$CO emission, the most common isotopologue of the most abundant molecule after H$_2$, allowing to trace the outer parts of disks. Most studies focused on the radial sizes in low- or mid- inclination disks 
which revealed rings, spirals, gaps \citep{Ansdell_2018, Avenhaus_2018, Pinte2018, Garufi2020, Galloway-Sprietsma_2025,Trapman_2025, Vioque_2025}. However, these disks do not allow the study of vertical sizes because of their viewing geometry. Highly inclined disks are the best objects for this study because their vertical extent is more directly accessible, with minimal de-projection required. Previous studies of edge-on disks at different wavelengths demonstrated that the concentration of pebbles is largely enhanced in the midplane \citep[e.g.,][]{Villenave2020, Villenave_2022, Duchene_2024, Tazaki_2025}. Moreover, highly inclined disks can provide better constraints on the disk outer radii because of the line-of-sight integration which compensates the low optical depth in the fainter outer regions. 

Beyond the disk structure and extent, the stellar properties are also fundamental to inform the planetary formation process. Indeed, several key disk parameters (e.g. radius, mass) are broadly correlated with the stellar mass \citep{Andrews2020}. Furthermore even the disk lifetime is correlated with the stellar mass \citep[e.g.][]{Ribas2015}. Unfortunately, it is not possible to determine the mass of the central star in highly inclined disks by comparison to evolutionary models in the HR diagram due to the lack of luminosity estimates. Indeed, the edge-on geometry of these systems blocks most of the stellar light, but exactly how much is extremely difficult to establish, the photosphere not being visible directly. If possible, estimating the dynamical mass thanks to the interaction between the host star and the gas of the disk in rotation is the most reliable method to determine this parameter independently of stellar models or viewing geometry \rev{\citep[e.g.,][]{Rosenfeld2012, Czekala2016, Simon_2019}. }

In this study, we present observations of a sample of highly inclined protoplanetary disks combining millimeter observations obtained with the Atacama Large Millimeter Array (ALMA) and optical/near-infrared (NIR) observations by the \textit{Hubble} Space Telescope (HST) and the ESO Very Large Telescope (VLT). This work is the follow-up of the study presented by \citet{Villenave2020}, with 5 new sources observed in the continuum and CO line data presented for the first time for all sources of the sample. We present estimations of the radial and vertical sizes of disks in $\mu$m-sized dust (scattered light), mm-sized dust (thermal continuum emission) and $^{12}$CO (3-2) gas line, as well as the dynamical masses of the central stars. 

The sample and the data reduction are described in Sect. \ref{sect:obs_and_data_red}. We present the images as well as our methods to extract the radial, vertical sizes and dynamical masses in Sect. \ref{sect:results_alma} for the ALMA continuum and $^{12}$CO images. Sect. \ref{sect:results_hst} shows the scattered-light HST and VLT images. After drawing comparisons on the distribution of dust of different sizes and $^{12}$CO gas, we discuss the differences in Sect. \ref{Sect:Discussion}, as well as trends with dynamical masses. Finally, we summarize our conclusions in Sect. \ref{sect:conclusions}.

\begin{table*}
\caption{Stellar parameters.}
\centering
\begin{tabular}{ccccccc}
\hline\hline
Full name & Adopted name & SFR & Distance & Spectral & Geometry & Multiplicity \\
&&&(pc)&type&\\
\hline
2MASS J16281370-2431391 & Flying Saucer   &  Ophiuchus & 140 & - & Edge-on & \\
IRAS 15462-2551         & PDS 144 N       &  Up Sco - Ophiuchus & 140 & A2$^{a}$ & Edge-on & Binary\\
2MASS J16230544-2302566 & Oph 162305      &  Ophiuchus & 140 & - & Edge-on &\\
2MASS J16070384-3911113 & Lup 160703      &  Lupus III & 160 & M4.5$^{b}$& Grazing-angle &\\
2MASS J16070854-3914075 & Lup 160708      &  Lupus III & 160 & M1.75$^{c}$& Grazing-angle & \\
\hline
HK Tau B                & HK Tau B        &  Taurus & 130 & M0.5$^{d}$ & Edge-on & Binary \\
HV Tau C                & HV Tau C        &  Taurus & 130 & K6$^{e}$ & Edge-on & Multiple \\
2MASS J04220069+2657324 & Haro 6-5B       &  Taurus & 130 & K5$^{e}$ & Grazing-angle &  \\
IRAS 04200+2759         & Tau 042307      &  Taurus & 130 & M3.5-M6$^{e}$ & Grazing-angle & \\
2MASS J04202144+2813491 & Tau 042021      &  Taurus & 130 & M1$^{f}$  & Edge-on & \\
ESO-H$\alpha$ 569       & ESO H$\alpha$ 569& Chamaeleon I & 190 & M2.5$^{g}$ & Edge-on & \\
ESO-H$\alpha$ 574       & ESO H$\alpha$ 574& Chamaeleon I & 190 & K7-M0$^{g}$ & Edge-on & \\
HH 48 NE                & HH 48 NE         & Chamaeleon I  & 190 & K7$^{g}$ & Edge-on & Binary \\
2MASS J16313124-2426281 & Oph 163131       & Ophiuchus & 140 & K4-K5$^{h}$ & Edge-on & \\
\hline
\label{table:stellar_parameters}
\end{tabular}
\tablefoot{SFR: Star-Forming Region. The distances come from \citet{Zucker_2020}. The horizontal line delimitates the new sources and the previous sample analyzed in the continuum by \citet{Villenave2020}.\\
\textbf{References.}
$^{a}${\citet{Vieira2003}},
$^{b}${\citet{Ansdell_2018}},
$^{c}${\citet{Muzic_2014}},
$^{d}${\citet{Monin_1998}},
$^{e}${\citet{Luhman_2010}},
$^{f}${\citet{Luhman2009}},
$^{g}${\citet{Luhman_2007}},
$^{h}${\citet{Flores_2021}}.
}
\end{table*}

\section{Observations and data reduction}
\label{sect:obs_and_data_red}


\subsection{Target selection}

This study is a follow up of the survey presented in \citet{Villenave2020} which combined several observing programs in Taurus, Ophiuchus and Chamaeleon (Projects 2013.1.01175.S, 2016.1.01505.S, 2016.1.00771.S, 2016.1.00460.S,
PIs: C. Dougados, F. Louvet, G. Duch\^ene and F. M\'enard). The authors analyzed a sample of 12 highly inclined disks in continuum emission. In this paper, we focus on 5 new highly inclined disks in Ophiuchus, Upper Scorpius and Lupus, in both millimeter continuuum and $^{12}$CO gas (Project 2022.1.00742.S, PI: F. M\'enard). CO data for 9 disks from the previous survey are also included. We choose to not include 3 disks from the previous survey
either because of the presence of an envelope, a CO outflow, or because of the circumbinary nature of the source (IRAS\,04302, HH\,30, and IRAS\,04158 respectively). Thus, our final sample consists of 14 highly inclined protoplanetary disks. 

All sources were identified as candidates from their spectral energy distribution and confirmed as highly inclined by optical and/or NIR scattered-light imaging \citep{Stapelfeldt_2014}. The properties of the sample in our study are reported in Table \ref{table:stellar_parameters}. We considered 3 categories in our sample: edge-on disks in single and multiple systems and, grazing angle disks. As their names indicate, the edge-on disks are the most inclined ones. They totally block the direct stellar light and only two bright reflection nebulae on both sides of the midplane are visible in scattered light. The grazing-angle disks are less inclined and partially mask the central star but a central point source is visible, with less pronounced nebulae in scattered light. The trends between the categories are discussed in Sect. \ref{Sect:Discussion}. 

\subsection{New ALMA data}


The new data are from Project 2022.1.00742.S (PI: F. M\'enard). The first scheduling block (SB) was executed on 6 June 2023 and included Lup\,160703 and Lup\,160708 ($\sim$18 minutes on source per target). The second SB was executed on 7 June 2023 for the Flying\,Saucer, PDS\,144\,N, and Oph\,162305 ($\sim$14 minutes on source per target). In both cases, 44 antennas were used during the Band 7 ($\sim$0.88~mm) observations, with baselines ranging from 27~m to $\sim$3.6~km. The correlator was set up with four spectral windows (SPW), three of which with a 1.875~GHz bandwidth (128~channels) targeting continuum (centered at 332.2, 334.0, and 344.5~GHz for the first SB, and at 333.0, 340.0, and 344.5~GHz for the second SB). The last SPW was set to a 0.469~GHz bandwidth (1920 channels, 0.2  km\,s$^{-1}$ per channel) targeting the $^{12}$CO (3-2) line (rest frequency 345.796~GHz).

The data were calibrated using the ALMA pipeline with {\tt CASA} version 6.5.4.9. For the Flying\,Saucer, PDS\,144\,N, and Oph\,162305, we performed between one and three rounds of phase-only self-calibration using the continuum emission to improve the quality of the observations, and the correspondings solutions were then applied to the $^{12}$CO (3-2) data. After subtracting the continuum emission, we imaged the $^{12}$CO (3-2) line using channels of $\sim$0.8 km\,s$^{-1}$. For the imaging, we used the {\tt tclean} task in {\tt CASA} with Briggs weighting and a robust parameter of 1.0. The final beam sizes of the $^{12}$CO (3-2) data cubes are listed in Table~\ref{table:beam_sizes}.

\subsection{Archival ALMA data}


The second part of this survey is based on ALMA observations from project 2016.1.00460.S~(PI: F. M\'enard), for which the continuum images have previously been published in \citet{Villenave2020}. In this program, three Cha~I sources (\ESOSevFour, \EsoSixNine, \hhq) have been observed in a compact array configuration, and 6 sources located in Taurus (\tauZero, \hk, \hv, \Haro, Tau\,042307) were observed with two array configurations (compact and extended), allowing to reach a higher angular resolution without suffering from significant flux losses. We refer the reader to \citet{Villenave2020} for a detailed description of the setup and calibration of the observations. In this study, we focus on the $^{12}$CO J=3-2 emission line of center frequency 345.796\,GHz. The intrinsic velocity resolution of the observations is 0.21 km/s. 

To increase the signal-to-noise and dynamical range of the observations, we self-calibrated the continuum visibilities of the brightest sources~(\tauZero, \hk, \hv, \Haro), and applied the continuum self-calibration solution to the line spectral window. Then, we extracted the continuum emission from the line spectral window using the \texttt{uvcontsub} CASA task. We imaged the CO emission line with the \texttt{tclean} task and the \texttt{multiscale} option, using a Briggs robust weighting parameter of 0.5, a velocity resolution of 0.25\,km/s. We use scales of 0, 1, 3, and 7 times the beam FWHM for all sources. 
In addition, for the very extended sources \Haro\ and \hv, we added one extra scale of respectively 12, 10, and 10 times the beam size. Finally, to improve the signal-to-noise ratio of the faintest targets~(namely \tauZero, \hk, \hv, \Haro, and Tau\,042307), we used the \texttt{uvtaper} option to increase the beam size. We report the final beam sizes in Table~\ref{table:beam_sizes}. 

We also include Band 6 $^{12}$CO (2-1) observations of Oph\,163131 from projects 2016.1.00771.S (PI: G. Duch\^ene) and 2018.1.00958.S (PI: M. Villenave). The observations combined a compact and an extended array configuration and were previously published in \cite{Villenave_2022}. We refer to that study for further description on the data reduction. 

\subsection{Scattered light archival data (HST and VLT)}

To compare the observed extent of our target disks between the submillimeter continuum and CO emission from ALMA to that observed in scattered light, we gathered HST optical and near-infrared Level 3 images from the Mikulski Archive for Space Telescopes (MAST). Specifically, we 
retrieved the NICMOS F110W image of the Flying\,Saucer and the ACS F814W image of PDS\,144\,N from program 10603 (PI: D. Padgett), the ACS F814W images of Lup\,160703 and Lup\,160708 from program 14212 (PI: K. Stapelfeldt), and the WFC3 F814W image of Oph\,162305 from program 17067 (PI: G. Duch\^ene). All archival images are fully calibrated. Some of these images were previously published \citep{
Stapelfeldt1998, Duchene_2010, Wolff2017, Wolff2021, Sturm2023_HH48_modeling, Duchene_2024}. 
In the case of HV Tau C, the HST optical images present a rather extended and round morphology indicating the presence of a remnant envelope. To better assess the size of the scattered light disk, we instead used the 2\,$\mu$m VLT/NaCo adaptive optics image presented in \citet{Duchene_2010}.


\section{Results from ALMA}
\label{sect:results_alma}

In this section, we estimate the radial and vertical extents of the millimeter continuum, and $^{12}$CO emission. For the systems studied in continuum in \citet{Villenave2020}, we simply report their results. 

\subsection{ALMA continuum images and $^{12}$CO emission moment maps}

\begin{figure*}[h!]
        \centering
        \includegraphics[width=1.0\textwidth]{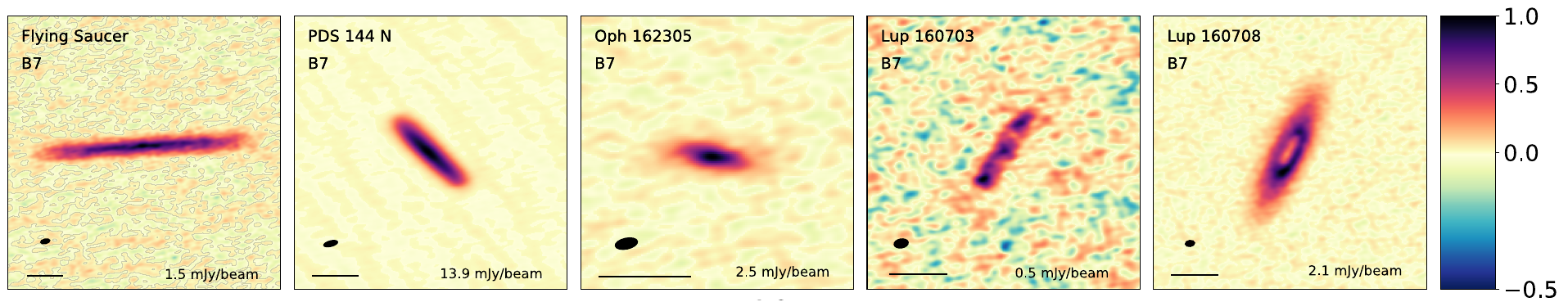}
        \caption{Continuum images of the newly observed sources
        normalized to their peak intensity (reported in the bottom right corner of each image). The beam size (ellipse) and a 0\farcs5 scale (dark line) are shown in the bottom left corner of each panel.}
        \label{fig:IMAGES_continuum}
\end{figure*}

All millimeter-wavelength continuum images of our new sample are presented in Fig. \ref{fig:IMAGES_continuum}. All of our disks present a continuous and very elongated emission shape with a large axis ratio, confirming their highly inclined geometry, except for Lup\,160703 and Lup\,160708 which show a knotty structure (possibly due to low signal-to-noise ratio, SNR) and a ring, respectively. These disks are less inclined than others, allowing us to observe structures that are not visible in the edge-on geometry. This geometry, which we refer to as grazing-angle, was also observed in the previous continuum survey for Haro\,6-5B and IRAS\,04158 \citep{Villenave2020}. We determined the position angle of each disks by manually measuring the orientation of their major axes (column 2 in Table \ref{table:DISKS_PARAMETERS}).

\begin{figure*}[h!]
        \centering
        \includegraphics[width=0.95\textwidth]{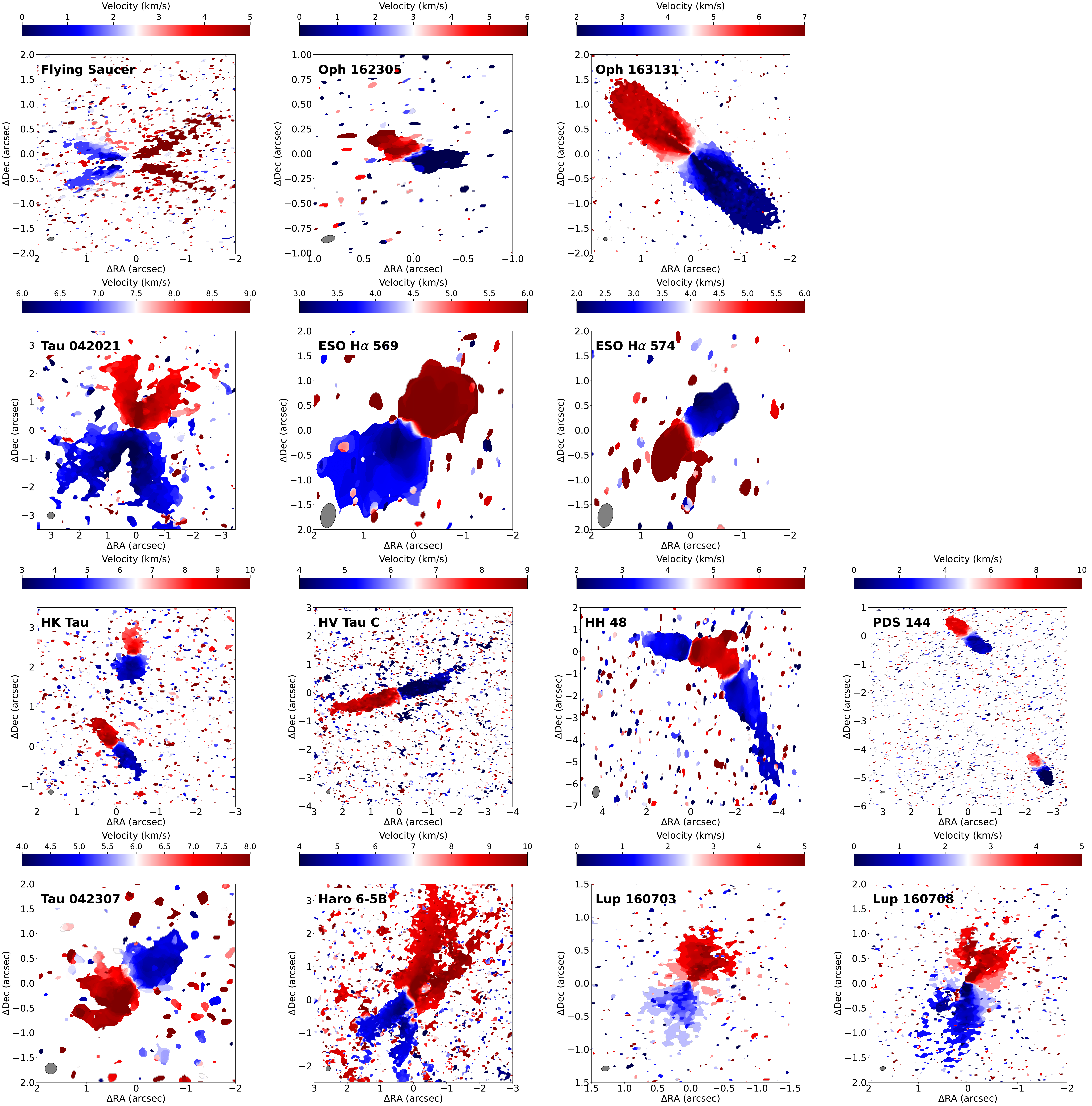}
        \caption{$^{12}$CO moment 1 maps of the 14 disks in both surveys. The first two rows correspond to edge-on disks, the third row to multiple systems with one edge-on disk and the last row to grazing-angle disks.}
        \label{fig:M1_MAPS_ALL}
\end{figure*}

The results of the $^{12}$CO (3-2) gas from both surveys are presented in Figs. \ref{fig:M1_MAPS_ALL} and \ref{fig:Overlays} for the moment 1 and moment 0 (M1 and M0) maps, respectively. We constructed the moment maps using the {\tt{bettermoments}} package developed by \citet{Teague_2018}, applying a 3$\sigma$ clipping. The companions in binary systems (third row on Fig. \ref{fig:M1_MAPS_ALL}) are also displayed in the moment maps, except for the HV\,Tau system for which the companions do not host a disk. We also notice that the gas emission of the two disks of the binary system HH\,48 overlap with each other. HH\,48\,SW presents a long tail (see Sect. \ref{sect:multiple_systems}) but we did not take it into account in the measure of the line flux.

We identified three different geometries in the moment maps of Fig. \ref{fig:M1_MAPS_ALL}: the typical rectangular elongated shape of edge-on disks (HK\,Tau\,B, HV\,Tau\,C, HH\,48\,NE, PDS\,144\,N -- intriguingly, all are members of multiple systems), the X-shape associated with edge-on disks   \citep{Dutrey_2017,Louvet_2018,Duchene_2024} with sharp edges (Flying\,Saucer, Tau\,042021, Oph\,163131), and finally, the fluffy X-shape for the less inclined disks (Haro\,6-5B, Lup\,160703 and Lup\,160708).  
The X-shape is due to the projection of CO emission layers in the warm molecular layer separated by a cold zone where CO freezes out. For the thinner disks, the two layers are not resolved leading to the rectangular shape. Regarding the grazing-angle disks (bottom row on Fig. \ref{fig:M1_MAPS_ALL}), their $^{12}$CO gas shape seems different from the vertically narrow emission or the sharp X-shape of edge-on disks, with a more spread-out emission. Their X-shape is more fluffy and filled in. There is also a vertical asymmetry (particularly pronounced for Haro\,6-5B and Lup\,160708) with fainter CO emission in the lower nebula as identified from the scattered light images, as expected from the vertical temperature gradient in the molecular layer.
We do not attribute a category for Oph\,162305, ESO\,H$\alpha$\,569, ESO\,H$\alpha$\,574 and Tau\,042307 because of the modest angular resolution for the first three disks and irregular morphology, respectively.

\begin{table}
\caption{Disks parameters.}
\centering
\begin{threeparttable}
\begin{tabular}{cccc}
\hline\hline
Sources & PA & $i_{\mathrm{AxisRatio}}$ & $i_{literature}$ \\
& (\degr )& (\degr) & (\degr) \\
\hline
Flying Saucer   & 93  & >86 & 87.0 $\pm$ 0.1$^{b}$ \\
PDS 144 N       & 45  & >81 & 83 $\pm$ 1$^{c}$  \\
Oph 162305      & 77  & >72 & - \\
Lup 160703      & 146 & >73 & - \\
Lup 160708      & 154 & 66  & - \\
HK Tau B        & 41  & >83$^{a}$ & 85 $\pm$ 1$^{d}$ \\
HV Tau C        & 108 & >80$^{a}$ & $83^{+\substack{1}}_{-\substack{2}}$$^{e}$ \\
Haro 6-5B       & 145 & 74$\pm$ 2$^{a}$ & $81^{+\substack{1}}_{-\substack{2}}$$^{f}$\\
Tau 042307      & 129 & 69$\pm$ 2$^{a}$ & - \\
Tau 042021      & 164  &  >85$^{a}$ & 88$^{g}$ \\
ESO H$\alpha$ 569 & 144 & >64$^{a}$ & 87.3 $\pm$ 0.3$^{h}$ \\
ESO H$\alpha$ 574 & 141 & >69$^{a}$ & - \\
HH 48 NE          & 75  & >68$^{a}$ & 88 $\pm$ 1$^{i}$ \\
Oph 163131        & 49  & >80$^{a}$ & 84$^{j}$ \\
\hline
\end{tabular}
\label{table:DISKS_PARAMETERS}
\begin{tablenotes}[flushleft]
\footnotesize
\item[]
\begin{minipage}{\linewidth} 
\tablefoot{The $i_{\mathrm{AxisRatio}}$ values are estimated from the ALMA continuum data.}
\textbf{References}.
$^{a}$\citet{Villenave2020}, 
$^{b}$\citet{Dutrey_2025}, 
$^{c}$\citet{Perrin_2006}, 
$^{d}$\citet{McCabe_2011}, 
$^{e}$\citet{Duchene_2010}, 
$^{f}$\citet{Kirchschlager_2016}, 
$^{g}$\citet{Duchene_2024}, 
$^{h}$\citet{Wolff2017}, 
$^{i}$\citet{Sturm2023_HH48_modeling}, 
$^{j}$\citet{Villenave_2022}.
\end{minipage}
\end{tablenotes}
\end{threeparttable}
\label{table:DISK_PARAMETERS}
\end{table}

\subsection{Millimeter continuum and $^{12}$CO (3-2) line fluxes}

\begin{table}[h]
\caption{ALMA Source Brightness Measurements}
\centering
\begin{threeparttable}
\begin{tabular}{cccc}
\hline\hline
Sources & Continuum & $^{12}$CO (3-2) & Central \\
& flux density & line flux & wavelength \\
& (mJy) &(Jy km s$^{-1}$) & ($\mu$m)\\
\hline
Flying Saucer   & 61 $\pm$ 6 & 1.2 $\pm$ 0.1 & 868 \\
PDS 144 N       & 213 $\pm$ 21 & 5.4 $\pm$ 0.5 & 868\\
Oph 162305      & 12 $\pm$ 1 & 0.43 $\pm$ 0.04 & 868 \\
Lup 160703      & 4.30 $\pm$ 0.40 & 1.3 $\pm$ 0.1 & 868\\
Lup 160708      & 76 $\pm$ 8 & 3.0 $\pm$ 0.3 & 868\\
HK Tau B        & 56 $\pm$ 6$^{a}$ & 1.5 $\pm$ 0.2 & 868\\
HV Tau C        & 91 $\pm$ 9$^{a}$ & 7.9 $\pm$ 0.8 & 868\\
Haro 6-5B       & 341 $\pm$ 34$^{a}$ & 9.7 $\pm$ 1.0 & 868\\
Tau 042307      & 66 $\pm$ 7$^{a}$ & 1.4 $\pm$ 0.1 & 868\\
Tau 042021      & 124 $\pm$ 12$^{a}$ &  6.8 $\pm$ 0.7 & 868\\
ESO H$\alpha$ 569 & 40 $\pm$ 4$^{a}$ & 2.3 $\pm$ 0.2 & 868\\
ESO H$\alpha$ 574 & 9 $\pm$ 1$^{a}$ & 1.3 $\pm$ 0.1 & 868\\
HH 48 NE          & 31 $\pm$ 3$^{a}$ & 5.9 $\pm$ 0.6 & 868\\
Oph 163131        & 125.8 $\pm$ 2.4$^{b}$ & 5.7 $\pm$ 0.6$^{\ast}$ & 1300\\
\hline
\end{tabular}
\begin{tablenotes}[flushleft]
\footnotesize
\item[]
\begin{minipage}{\linewidth} 
\tablefoot{$^{\ast}$$^{12}$CO (2-1) line flux.}
\textbf{References}.
$^{a}$\citet{Villenave2020}, 
$^{b}$\citet{Cox_2017}.
\end{minipage}
\end{tablenotes}
\end{threeparttable}
\label{table:FLUX}
\end{table}

All continuum and gas brightness values are reported in Table \ref{table:FLUX}. The fluxes were measured by integrating within rectangular apertures tailored to each disk. We quadratically combined the measurement error and the calibration error estimated at 10$\%$ which correspond to the typical flux calibration errors of ALMA (see ALMA Technical Handbook\footnote{\url{https://almascience.eso.org/documents-and-tools/cycle12/alma-technical-handbook}}). We note that the error was dominated by the calibration one. The gas flux was extracted from the M0 maps, following the same process as the continuum flux with a 10$\%$ calibration error. 

\subsection{ALMA radial extents}

To measure the radial extent of the disks from their continuum emission we extracted surface brightness profiles along the major axis of each disk. We integrated over the full vertical extent of the disk and normalized by the maximum intensity. We then estimated the  noise level ($\sigma$) outside of the disk signal. 
The major axis size was determined at a level of 3$\sigma$ above the background. We have re-evaluated the uncertainties associated with the continuum sizes measurements in \citet{Villenave2020}, which previously accounted only for the beam-related uncertainty. The dominant source of error is methodological: the estimated size depends slightly  on the size over which we integrated to do the profile. Therefore, we estimated the associated uncertainty by varying the window over which the profile is estimated and quadratically added a tenth of the beam size along the major axis direction. 
Because of the vertical integration, this method is sensitive to lower signal which is not visible in contour images. The resulting major axis sizes are presented in column 2 of Table \ref{table:SIZES_MAJOR}.

The gas extent is more difficult to measure because the emission is intrinsically fuzzier and noisier than in the continuum, especially at the outer radii.
We measured the radial extent of the $^{12}$CO emission with cuts along the major axis directly through the image cubes. Instead of processing all the channels for which the $^{12}$CO emission is detected, we visually selected the channel for which the radial extent is maximum, for the blue and red side to maximize SNR. We then followed the same method described above for the size of the continuum emission to measure the radius at a level of 3$\sigma$ above the background. However, for HH\,48\,NE, on the West side the signal does not go down to 3$\sigma$ because of contamination by the companion. In this case, we estimated the maximal extent as the point of the minimum emission between the two disks, which slightly underestimates the size since this minimum is at the 5$\sigma$ level. The uncertainties are estimated following the same process as that used for the radial extent of the continuum.
The values are reported in column 4 of Table \ref{table:SIZES_MAJOR}. We note that \finalfinal{a few} of these values are larger by up to 40$\%$ \final{(in the most pathological case)} than those estimated from the moment \rev{0} maps \final{due to differences in noise statistics} \finalfinal{and cloud contamination,}  confirming the inadequacy of moment maps to estimate the full radial extent of disks. However, there is an exception for Flying\,Saucer, for which the value is lower compared to what is apparent in the moment map (Fig. \ref{fig:Overlays}) likely due to low SNR. For this disk, in the following part of our study, we adopt the diameter radial extent of 5$\farcs$5 \citep{Dutrey_2017}. 


\begin{table}
\caption{Major axis diameters measurements from the ALMA continuum, HST scattered light observations and ALMA $^{12}$CO.}
\centering
\begin{threeparttable}
\begin{tabular}{cccccccc}
\hline\hline
Sources & Millimeter & Optical/NIR &  Millimeter  \\
& continuum & scattered & $^{12}$CO gas\\
  & (\arcsec) & (\arcsec) & (\arcsec) \\
\hline
Flying Saucer   & 3.18 $\pm$ 0.05 & 3.2 $\pm$ 0.10 & 2.64 $\pm$ 0.10$^{\ast}$ \\
PDS 144 N       & 1.25 $\pm$ 0.05 & 1.7 $\pm$ 0.10 & 1.84 $\pm$ 0.10 \\
Oph 162305      & 0.67 $\pm$ 0.05 & 0.9 $\pm$ 0.10 & 1.02 $\pm$ 0.10  \\
Lup 160703      & 0.84 $\pm$ 0.05 & 1.7 $\pm$ 0.10 & 2.08 $\pm$ 0.10 \\
Lup 160708      & 1.78 $\pm$ 0.05 & 2.0 $\pm$ 0.10 & 3.04 $\pm$ 0.10 \\
HK Tau B        & 0.99 $\pm$ 0.05$^{a}$ & 1.3 $\pm$ 0.10$^{a}$ & 1.73 $\pm$ 0.10 \\
HV Tau C        & 1.20 $\pm$ 0.05$^{a}$ & 0.8 $\pm$ 0.10$^{a}$ & 3.74 $\pm$ 0.10  \\
Haro 6-5B       & 2.06 $\pm$ 0.05$^{a}$ & 3.4 $\pm$ 0.10 & 6.27 $\pm$ 0.10 \\
Tau 042307      & 1.00 $\pm$ 0.05$^{a}$ & 1.5 $\pm$ 0.10 & 2.49 $\pm$ 0.10\\
Tau 042021      & 4.10 $\pm$ 0.05$^{a}$ & 5.0 $\pm$ 0.10$^{a}$& 7.05 $\pm$ 0.10 \\
ESO H$\alpha$ 569 & 1.88 $\pm$ 0.06$^{a}$ & 2.0 $\pm$ 0.10$^{a}$ & 3.67 $\pm$ 0.11\\
ESO H$\alpha$ 574 & 1.35 $\pm$ 0.06$^{a}$  & 1.2 $\pm$ 0.10$^{a}$ & 2.32 $\pm$ 0.11 \\
HH 48 NE        & 1.72 $\pm$ 0.05$^{a}$ & 1.3 $\pm$ 0.10$^{a}$ & 3.40 $\pm$ 0.10\\
Oph 163131      & 2.50 $\pm$  0.05$^{a}$ & 3.0 $\pm$ 0.10 &  4.44 $\pm$ 0.10 \\
\hline
\end{tabular}
\begin{tablenotes}[flushleft]
\footnotesize
\item[]
\begin{minipage}{\linewidth} 
\tablefoot{$^{\ast}$The larger size of the Flying source in scattered light compared to the millimeter emission is likely due to the low SNR of the ALMA data. In the following part of our study, we used the diameter radial extent of 5$\farcs$5 \citep{Dutrey_2017}}
\textbf{References}.
$^{a}$\citet{Villenave2020}.
All observations are in Band 7 except for Oph\,163131 which are in Band 6.
\end{minipage}
\end{tablenotes}
\end{threeparttable}
\label{table:SIZES_MAJOR}
\end{table}

\subsection{ALMA vertical extents}
\label{sect:results_alma_vertical}
To measure the vertical extent of the disks from the continuum images, we extracted brightness profiles along the minor axis of each disk. We applied the same method as for the continuum radial extent. 
The profiles being single peaked because the continuum vertical extent is not well resolved, we estimated the FWHM by fitting Gaussians to the disk profiles 
except for Lup\,160708 where the ring geometry prevents a measurement of the intrinsic vertical extent. 
The measurements errors correspond to a tenth of the beam size projected in the direction of the cut. We also quadratically subtract the FWHM of the beam from the measured FWHM to obtain the intrinsic FWHM and propagated the errors.
The resulting minor axis sizes are presented in column 2 in Table \ref{table:SIZES_MINOR}.

We also estimated the disk inclinations from their measured axis ratio. To measure the axis ratio, we use the major axis size at 50$\%$ of the maximum flux that can be readily compared to the intrinsic FWHM along the minor axis. The values are reported in column 3 in Table \ref{table:DISKS_PARAMETERS} as lower limits because the aspect ratio is a combination of the intrinsic vertical extent and projection effects, except for Lup\,160708 where an ellipse is fitted to the ring.

The vertical extent of the $^{12}$CO emission was also measured with cuts along the minor axis. Contrary to the radial extent, the vertical extent does not vary much from channel to channel.
We have chosen to collapse all channels for which the gas emission was detected to gain SNR. To estimate the gas vertical extent, we followed the same method as for the radial extent of the gas without fitting Gaussians to the disk profiles as for the continuum vertical extent because the gas is vertically well resolved compared to the continuum. In this case, the dominant uncertainty is the same described as for the radial extent of the continuum. We quadratically added 10\% of the beam size in the direction of the cut to that uncertainty.
The values are reported in column 4 in Table \ref{table:SIZES_MINOR}. 


\begin{table}
\caption{Minor axis sizes measurements from the ALMA continuum, HST scattered light observations and ALMA $^{12}$CO.}
\centering
\begin{threeparttable}
\begin{tabular}{cccc}
\hline\hline
Sources & Millimeter & Optical/NIR & Millimeter\\
&  continuum & scattered & $^{12}$CO gas\\
&  (\arcsec) & (\arcsec) & (\arcsec)\\
\hline
Flying Saucer   & 0.20 $\pm$ 0.01 & 1.13 $\pm$ 0.02 & 0.86 $\pm$  0.02$^{\ast}$\\
PDS 144 N       &  0.14 $\pm$ 0.03  & 0.62 $\pm$ 0.02 & 0.60 $\pm$ 0.03\\
Oph 162305      & 0.10 $\pm$ 0.01 & 0.17 $\pm$ 0.03 & 0.28 $\pm$ 0.02 \\
Lup 160703      &  0.15 $\pm$ 0.02 & 0.35 $\pm$ 0.05 & 0.81 $\pm$ 0.02\\
Lup 160708      &  - & 0.65 $\pm$ 0.05 & 1.05 $\pm$ 0.02\\
HK Tau B        &  0.10 $\pm$ 0.01$^{a}$ & 0.23 $\pm$ 0.04 & 0.53 $\pm$ 0.02\\
HV Tau C        &  0.13 $\pm$ 0.02$^{a}$ & 0.30 $\pm$ 0.04 & 0.65  $\pm$ 0.02\\
Haro 6-5B       &  0.34 $\pm$  0.01$^{a}$ & 1.36 $\pm$ 0.02 & 1.33 $\pm$ 0.02\\
Tau 042307      &  0.31 $\pm$ 0.01$^{a}$ & 0.43 $\pm$ 0.03 & 1.03 $\pm$ 0.03\\
Tau 042021      & 0.31 $\pm$ 0.01$^{a}$ & 2.15 $\pm$ 0.05 & 2.18 $\pm$ 0.04\\
ESO H$\alpha$ 569 & <0.13$^{a}$ & 0.90 $\pm$ 0.02 & 1.06 $\pm$ 0.03\\
ESO H$\alpha$ 574 &  <0.21$^{a}$ & 0.40 $\pm$ 0.08 & 0.75 $\pm$ 0.04\\
HH 48 NE          &  <0.24$^{a}$ &0.40  $\pm$ 0.10 & 1.41 $\pm$ 0.04\\
Oph 163131        &  <0.14$^{a}$ &0.42 $\pm$ 0.06 & 0.97 $\pm$  0.02\\
\hline
\end{tabular}
\tablefoot{See Sect. \ref{sect:results_alma_vertical} and \ref{sect:results_hst_vertical} for definitions of the three quantities. 

$^{\ast}$The larger size of the Flying source in scattered light compared to the millimeter emission is likely due to the low SNR of the ALMA data.}
\begin{tablenotes}[flushleft]
\footnotesize
\item[]
\begin{minipage}{\linewidth} 
\textbf{References}.
$^{a}$\citet{Villenave2020}. All observations are in Band 7 except for Oph\,163131 which are in Band 6.
\end{minipage}
\end{tablenotes}
\end{threeparttable}
\label{table:SIZES_MINOR}
\end{table}

\subsection{Dynamical masses}
\label{sect:dynamical_masses}

\begin{table}
\caption{Systemic velocities and dynamical masses from this work and from the literature.} 
\centering
\begin{threeparttable}
\begin{tabular}{cccc}
\hline\hline
Sources & V$_{sys}$ & M$_{\star}$ & M$_{\star}$$_{literature}$ \\
  & (km s$^{-1}$) &  (M$\odot$) & (M$\odot$)\\
\hline
Flying Saucer & 3.2 & 0.9 (0.6..1.5) & 0.60$^{a}$\\
PDS 144 N  & 4.9 & 2.1 (1.4..2.8) & \\
Oph 162305  & 2.5 & 1.2 (0.7..1.7) & \\
Lup 160703  & 3.1 & 0.4 (0.2..0.6) & \\
Lup 160708  & 2.7 & 1.0 (0.5..1.3) & \\
HK Tau B    & 6.9 & 1.2 (0.9..1.5) & 0.89 $\pm$ 0.04$^{b}$\\
HV Tau C    & 6.4 & 1.9 (1.2..2.3) & 1.33 $\pm$ 0.04$^{b}$ \\
Haro 6-5B   & 7.5 & 1.6 (1.0..2.5) & 0.25$^{c}$ \\
Tau 042307  & 6.3 & 0.7 (0.5..1.0) & 0.52 $\pm$ 0.04$^{b}$\\
Tau 042021  & 7.4 & 0.5 (0.3..0.6) & 0.25 - 0.4$^{d}$\\
ESO H$\alpha$ 569 & 4.7 & 0.7 (0.4..0.9)\\
ESO H$\alpha$ 574 & 4.3 & 1.2 (0.7..1.5) \\
HH 48 NE    & 4.75 & 1.5 (1.0..2.5) & 1 - 1.4$^{e}$\\
Oph 163131  & 4.5 & 1.6 (1.1..2.2) & 1.2 $\pm$ 0.2$^{f}$\\
\hline
\end{tabular}
\tablefoot{We present a preferential mass with lower and upper limits.}
\begin{tablenotes}[flushleft]
\footnotesize
\item[]
\begin{minipage}{\linewidth} 
\textbf{References}.
$^{a}$\citet{Dutrey_2025},
$^{b}$\citet{Simon_2019}, 
$^{c}$\citet{Yokogawa_2002}, 
$^{d}$\citet{Duchene_2024}, 
$^{e}$\citet{Sturm2023_HH48_modeling}, 
$^{f}$\citet{Flores_2021}.
\end{minipage}
\end{tablenotes}
\end{threeparttable}
\label{table:MASSES}
\end{table}

\begin{figure*}[h!]
        \centering
        \includegraphics[width=1.1\textwidth, angle=-90]{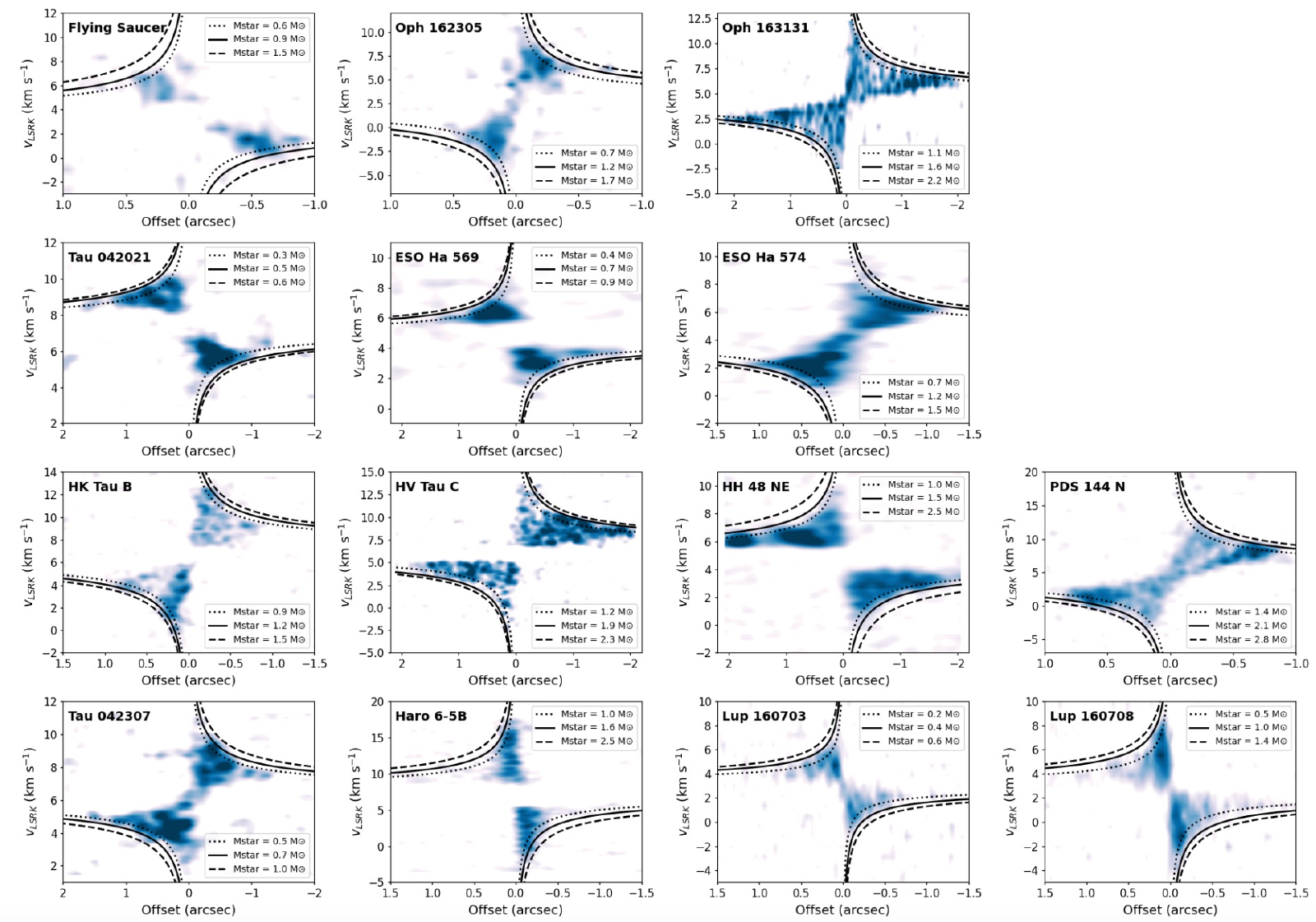}
        \caption{Position-velocity diagrams with 3 Keplerian velocity curves corresponding to the preferred dynamical mass of the star, as well as conservative upper and lower limits.}
        \label{fig:PV_DIAGRAMS_ALL}
\end{figure*}

Fig. \ref{fig:PV_DIAGRAMS_ALL} presents a position-velocity (PV) diagram for each source. Relevant keplerian velocity curves are superimposed in each diagram to estimate the stellar masses. To draw these curves, we considered inclinations from the literature when available or used the results from column 3 of Table \ref{table:DISKS_PARAMETERS} for systems with no previous values. For the latter, when a lower limit is estimated, we considered the midpoint between this lower limit and 90°, but note that the PV diagram is hardly affected by the inclination in nearly edge-on configurations. \rev{Since GAIA DR3 distance measurements are unreliable for edge-on disks,} we considered distance of 130\,pc for Taurus, 140\,pc for Ophiuchus, 140\,pc for Upper-Scorpius, 160\,pc for Lupus and 190\,pc for Chamaeleon \citep{Zucker_2020}. \rev{We note, however, that the dynamical masses scale linearly with distance.} We note that two disks only contain information along the vertical direction in their PV diagrams (Haro\,6-5B and Lup\,160708). This can be the tracer of a ring geometry \citep{Dutrey_2017}, which is confirmed in the millimeter continuum image of Lup\,160708 (see Fig. \ref{fig:IMAGES_continuum}) and Haro\,6-5B \citep{Villenave2020}.

The estimation of stellar masses from PV diagrams is \rev{generally ill defined due to the modest SNR per spaxel, cloud contamination, and the presence of possible substructures and asymmetries, which precludes a robust statistical approach.} 
With a coarse exploration, we thus choose to add 3 Keplerian velocity curves: a first one which fits the outermost maximum intensity of the PV diagram, a second one which follows the transition between the signal and the noise and a last one clearly off the signal. These 3 Keplerian curves allow to constrain a lower limit, a preferred mass as well as an upper limit, respectively. The lower and upper limits should be treated as very conservative ($\approx3\sigma$) limits. We visually estimated the systemic velocity so that the Keplerian curves were symmetrically aligned with the PV diagrams.
All values of systemic velocity and stellar mass are summarized in Table. \ref{table:MASSES}. 

We also notice that the lowest dynamical masses (ranging from 0.5 to 1.0 M$_\odot$) correspond to 3 out of 4 grazing-angle systems (Tau\,042307, Lup\,160703, Lup\,160708) and 3 edge-on systems (Flying\,Saucer, ESO\,H$\alpha$\,569, Tau\,042021). Furthermore, 4 disks out of 6 among the lowest masses show a clear X-shape. 
A possible interpretation is that a low dynamical mass implies a larger $h/r$ ratio (Fig. \ref{fig:mass_vs_hr}), which is a case where the X-shape is well-defined if the angular resolution is sufficiently high. 

\section{Results from HST and VLT optical and near-infrared scattered light}
\label{sect:results_hst}

In this section, we estimate the radial and vertical extents of the scattered light. \citet{Villenave2020} reported the radial extent of all but one of their disks and we report their results here. We also determine the radial extent of Tau\,042307, which they had not estimated.

\subsection{Optical and NIR scattered light images}

Fig. \ref{fig:Overlays} shows overlays of optical and NIR scattered light images with continuum millimeter and $^{12}$CO images. The scattered light images in the first three rows present the typical features of edge-on disks: two bright reflection nebulae separated by a dark lane blocking direct starlight. However, the morphology of disks shown in the last row is different with a less visible bottom nebula and a visible central point source. These disks are seen at grazing angle (i.e. inclined so that the stellar light lay in our line-of-sight).
We notice a bizarre observation for Oph\,162305. Indeed, the PA in the HST 
scattered light image (69\degr) does not match the PA in the millimeter continuum image (77\degr). A possible explanation lies in the fact that, Oph\,162305 is a very compact disk resulting in less well defined PAs. Moreover, the ALMA beam is tilted relative to the disk major axis and may affect the interpretation. 
Higher quality ALMA observations are needed to conclude whether this is a physical effect or an artifact of the observations.

\subsection{Optical and NIR scattered light radial extent}

We estimated the scattered light major axis sizes by following the spine of the more extended nebula, based on the method described in Appendix D in \citet{Villenave2020}. The errors correspond to a typical value of $\sim$2 pixels in the image due to the uncertain transition between clearly detected signal and background noise. We checked the radial extents of the disks from the previous survey and updated them for Haro\,6-5B and Oph\,163131, which had been underestimated. The major axis sizes are summarized in column 3 of Table \ref{table:SIZES_MAJOR}.

\subsection{Optical and NIR scattered light vertical extent}
\label{sect:results_hst_vertical}

We also estimated the scattered light minor axis sizes based on the spine method used to estimate the major sizes for all the sources. In this case, we considered the distance along the vertical direction between the upper and lower nebulae at the edges of the disk. The vertical extent is defined as the mean between the vertical extent on both sides of the disk. The errors correspond to half the difference between these two values. The values are summarized in column 3 of Table \ref{table:SIZES_MINOR}. However, these sizes should be considered as underestimates if the disk radially extends further on one side than the other.

\begin{figure*}[h!]
        \centering
        \includegraphics[width=1\textwidth]{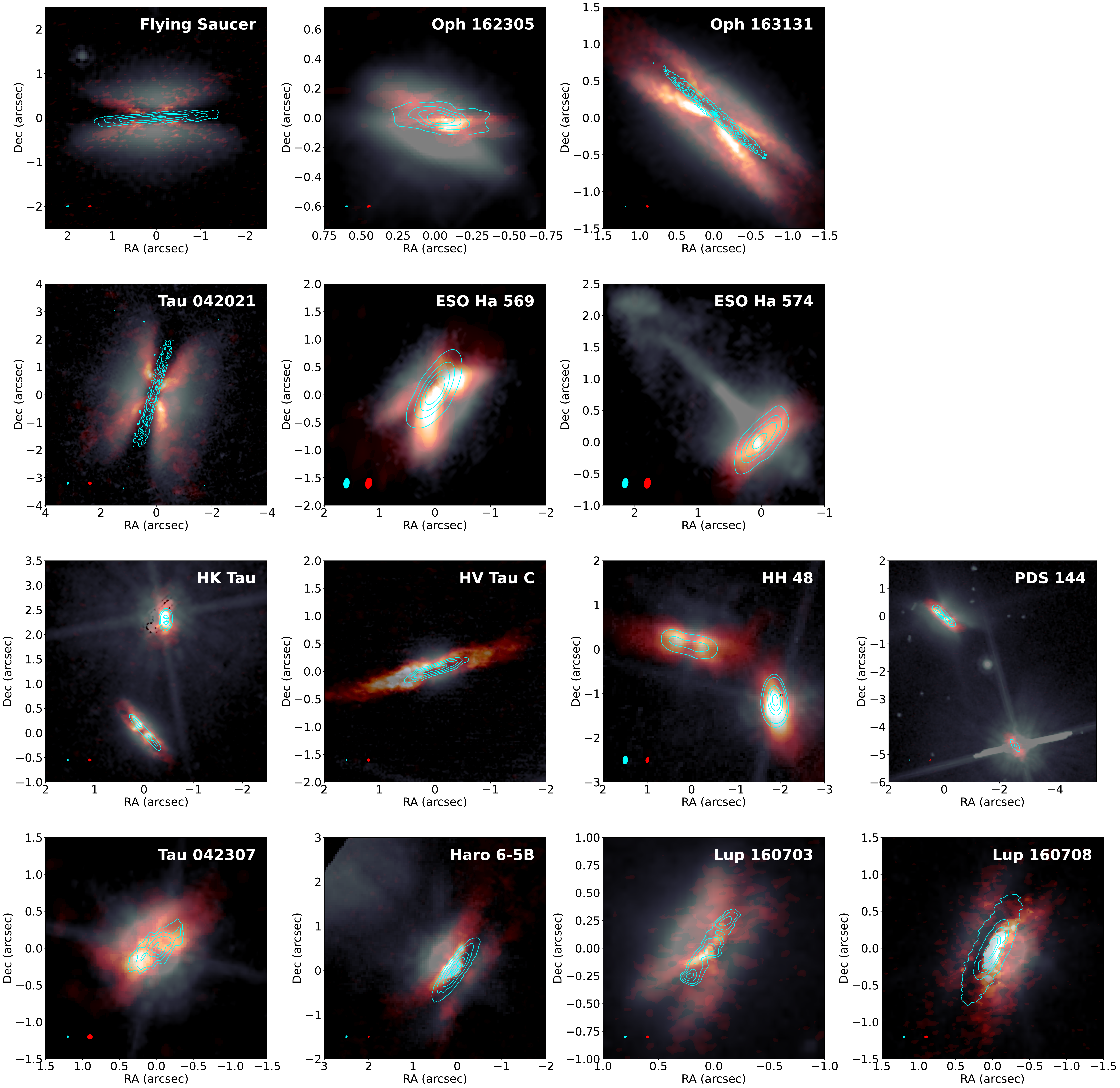}
        \caption{Overlay of scattered light (grayscale image), ALMA 0.9\,mm continuum (blue contours) and ALMA $^{12}$CO gas (semi-transparent red image) for all sources in this study. The scattered light images are plotted on a logarithmic stretch and the contours of ALMA continuum are plotted at 30, 50, 70, and 90$\%$ of the maximum emission, except for the binaries for which we adjusted the contours due to the brightness of the companion as well as for Oph\,163131 to distinguish the rings structure. The blue and red ellipses in the bottom left corner indicate the beam size of the ALMA continuum and ALMA $^{12}$CO gas, respectively. Scattered light images: Flying\,Saucer: 1.1\,\mic ; Oph\,162305: 0.81\,\mic ; Oph\,163131: 0.81\,\mic ; Tau\,042021: 0.81\,\mic ; ESO\,H$\alpha$\,569: 0.81\,\mic ; ESO\,H$\alpha$\,574: 0.6\,\mic ; HK\,Tau: 0.48\,\mic ; HV\,Tau\,C: 2.1\,\mic ; HH\,48: 0.81\,\mic ; PDS\,144: 0.81\,\mic ;  Tau\,042307: 0.6\,\mic ; Haro\,6-5B: 1.6\,\mic ; Lup\,160703: 0.81\,\mic ; Lup\,160708: 0.8\,\mic.}
        \label{fig:Overlays}
\end{figure*}

\section{Discussion}
\label{Sect:Discussion}

\subsection{Comparison of radial and vertical disk sizes}
\label{sect:comparison_sizes}

\begin{figure*}[h!]
        \centering
        \includegraphics[width=0.9\textwidth]{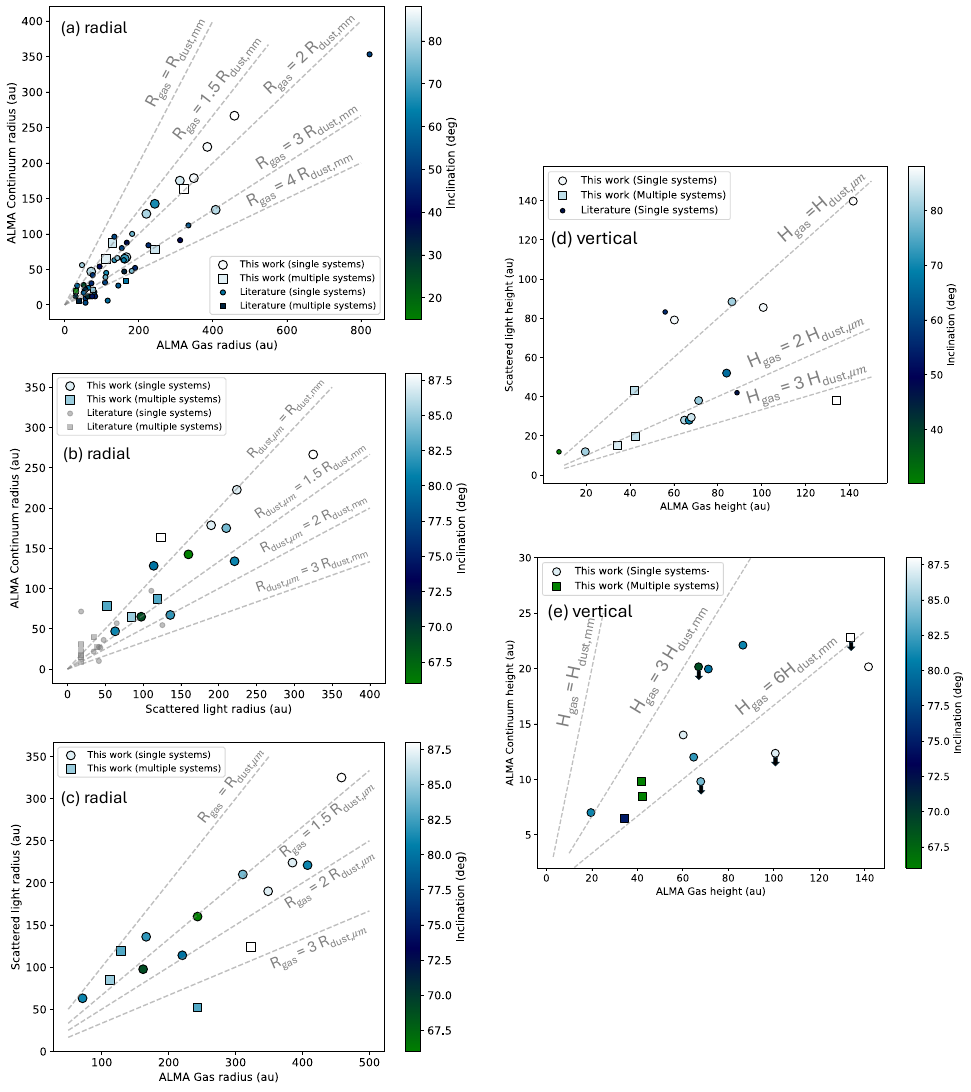}
        \caption{Comparison between disks radius in gas, millimeter dust and scattered light dust in the radial direction (top row) and disks heights (bottom row). We added radius measurements of Class II disks in the millimeter dust from \citet{Ansdell_2018, Vioque_2025}, in the scattered light dust from \citet{Avenhaus_2018, Garufi2020}
        and in the gas from \citet{Pinte2018, Galloway-Sprietsma_2025}. The inclinations are not indicated in the sample of \citet{Garufi2020}, however, all the disks are at low inclinations, except one disk which is the more extended in millimeter and scattered light.}
        \label{fig:discuss_sizes}
\end{figure*}

The differences in radial and vertical extents between gas, millimeter dust and scattered light dust allow us to draw interpretations related to the growth and dynamics of grains. In this subsection, we compare the radius size, $R$ (left column of Fig. \ref{fig:discuss_sizes}), and the vertical apparent emitting height, $H$, --which should not be confused with the scale height-- (right column of Fig. \ref{fig:discuss_sizes}) between the gas and the two population of dust sizes. In the following, we suppose that the $^{12}$CO emission and the $\mu$m-sized dust are much more optically thick than the millimeter continuum. 

All disks in our new sample are more radially extended in scattered light than in the millimeter continuum (see Fig. \ref{fig:Overlays}). 
The conclusion is similar for the disks of the previous survey, with the exception of HV\,Tau\,C, ESO\,H$\alpha$\,574 and HH\,48\,NE that are more compact in scattered light. Two of these three disks are members of multiple systems and we will discuss differences in these systems in Sect. \ref{sect:multiple_systems}. 
The more extended scattered light compared to the continuum millimeter can be linked to dust radial drift that drags dust grains towards pressure maximum or opacity effects. Moreover, all disks are more extended in the $^{12}$CO gas emission than in scattered light and in the continuum emission. These results are discused in Sect.\,\ref{Sect:Discussion_radial_extent}.

As for the radial extent, the disks appear clearly vertically extended in scattered light as opposed to the millimeter continuum which appears flat and concentrated to the midplane. 
On the other hand, the vertical extent of the $^{12}$CO gas is closer to the vertical extent of the scattered light images. 
This is expected since the scattered light images probe the small \mic-sized grains that are expected to be well coupled to the gas, whereas the millimeter continuum probes bigger grains that have settled to the midplane \citep{Villenave2020}. This is further discused in Sect.\,\ref{Sect:Discussion_vertical_extent}.

\subsubsection{Radial extents} 
\label{Sect:Discussion_radial_extent}

In all three tracers (millimeter continuum, CO emission and scattered light), highly inclined disks appear generally larger than lower inclination ones. This is a consequence of a known selection bias as only the larger edge-on disks are well-resolved in the HST images used to identify them in the first place \citep{Angelo2023}.

\textit{Continuum emission vs gas.} To better place edge-on disks in context,
we also added size measurements of less inclined Class II disks from the literature \citep{Ansdell_2018, Trapman_2025, Vioque_2025}, see Fig.~\ref{fig:discuss_sizes}a. The disks in our sample follow the trend of less inclined disks with $R_\mathrm{gas}$ larger than $R_\mathrm{dust, \, mm}$. For the general population, the $R_\mathrm{gas}$ / $R_\mathrm{dust, \, mm}$  ratio is centered around 2--3 \citep{Ansdell_2018, Deng_2025}. However, the highly inclined disks tend to have smaller gas-to-dust radius ratios, and appear closer to $R_\mathrm{gas}$
$\approx$ 1.5--2 $\times$ $R_\mathrm{dust, \, mm}$. These smaller ratios may be expected because, e.g.,  of line-of-sight integration effects, the highly inclined disks providing more robust estimates of $R_\mathrm{dust, \, mm}$, with larger values than for less inclined disks.
Is it worth noting however that statistical tests show that the difference between the two samples is marginal. The Mann-Whitney $U$ test, most sensitive to the median value of each sample, yields a $\approx$90\% confidence level that the two populations are intrinsically different. Similarly, the Kolmogorov-Smirnov test, which is more sensitive to the shape of the underlying distribution, yields a 93\% confidence level. 

The smaller dust radius compared to the gas radius can be explained in more details by a combination of optical depth effects and radial drift \citep{Facchini_2017, Ansdell_2018}. Indeed, the millimeter dust is less optically thick than the $^{12}$CO emission, which can lead to an easier detection of gas at larger radii. This effect is attenuated in the case of highly inclined disks, where line-of-sight integration enhances the detectability of the continuum emission in the outer regions.
From a model perspective, \citet{Ansdell_2018}  and \citet{Trapman_2019} showed that a $R_\mathrm{gas}$ / $R_\mathrm{dust, \, mm}$ ratio $\gtrsim$ 3--4 is a clear indicator of grain growth and radial drift. Conversely, \citet{Ansdell_2018}, following the models of \citet{Facchini_2017}, found $R_\mathrm{gas}$ / $R_\mathrm{dust, \, mm}$ $\approx$ 1.5 for a model with uniform dust distribution, i.e., no growth and no drift. Only a few disks in the literature sample and none of our highly inclined disks sample are above the maximum ratio predicted by the radial drift models (see Fig. \ref{fig:discuss_sizes}a). Taken at face value, these results suggest that the ratios in most of our highly inclined disks can be explained by optical depth effects because their $R_\mathrm{gas}$ / $R_\mathrm{dust, \, mm}$ ratios are close to 1.5. However, these model ratios are predicted in the case of mid-inclinations disks and do not account for (large) radiative transfer effects expected in very highly inclined ones. Models of radial drift with radiative transfer dedicated to highly inclined disks are required to disentangle between radial drift and opacity effects based on the $R_\mathrm{gas}$ / $R_\mathrm{dust, \, mm}$ ratio.

\textit{Continuum emission vs scattered light.}
The comparison between the millimeter (large grains) and scattered light (small grains) dust is more poorly constrained in the literature. We added disk size estimates of all objects from \citet{Garufi2020} to our sample. The scattered light dust is more radially extended than the millimeter dust, except for 4 disks (see Fig. \ref{fig:discuss_sizes}b). The disks tend to follow $R_\mathrm{dust,\,\mu m} / R_\mathrm{dust, \, mm}$ $\approx$ 1--2, which is consistent with the results in \citet{Villenave2020} and with the radial drift theory between millimeter and scattered light dust \citep[$R_\mathrm{dust,\,\mu m}$/$R_\mathrm{dust, \, mm}$ $\approx$ 1.6; e.g.,][]{Laibe_2014}. In other words, the disks with clear indicator of radial drift are those for which the millimeter dust is less extended than the scattered light dust. {\rev{Ultimately, this ratio can be an indicator of more or less evolved disks, although it should not be used as the single argument} \final{Ultimately, this ratio may represent a potentially more reliable indicator of dust radial drift, a claim that dedicated models should verify. One must keep in mind however that the more evolved disks, with traces of more radial drift, are not necessarily the older ones.}


\textit{Scattered light vs gas.} Regarding the comparison between the scattered light (small grains) dust and the gas, the disks are more radially extended in gas than in scattered light (see Fig. \ref{fig:discuss_sizes}c). All the disks in our sample are in the range  $R_\mathrm{gas} / R_\mathrm{dust,\,\mu m}$ $\approx$ 1--2 , except two disks, which are in multiple systems, and have a size ratio above 2 and 3. This could be explained by the fact that the dust in the outer regions is not illuminated by the central star, due to shadowing, while the CO emission is still detectable, or there are not enough $\mu$m-sized grains to be detected. Nonetheless, several indications suggest that only a small amount of $\mu$m dust lays in the midplane in the outermost regions probed by the scattered light, such as the silhouette in Flying\,Saucer \citep[][ Ma et al. in prep]{Dartois2024} and the detection of a galaxy behind Tau\,042021 \citep{Duchene_2024}. Another explanation suggests that the small dust is not perfectly radially coupled to the gas which extends over larger radii. This is in line with the analytical models  \citep{Birnstiel2014}, who demonstrated that even $\mu$m-sized dust can drift inward in the outermost regions of a disk, where the gas density is low enough. These models suggest that most protoplanetary disks should present a sharp outer edge in their dust component well inside the outer gas radius, consistent with our findings. \citet{Barriere_Fouchet_2005} also showed in models without dust evolution that the outer radius of $\mu$m-sized dust is smaller than the gas outer radius.

Strong correlations between flux density and disk radius have been documented in the literature for both the continuum \citep[e.g.,][]{Tripathi2017, Andrews_2018, Hendler_2020} \rev{and line emission \citep[e.g.,][]{Sanchis_2021, Long2022, Zagaria_2023, Zallio_2026}}. This is also observed in our sample of highly inclined disks, where the two quantities are significantly correlated ($r=0.47$ \rev{and $r=0.67$}) with confidence levels of 91\% \rev{and 99.2\%} based on Pearson tests \rev{for the continuum and line emission, respectively}. Since the flux density roughly scales as $R_\mathrm{dust, \, mm}^2$, the most common interpretation of this correlation is a geometrical effect that arises if disks are optically thick and the total flux primarily depends on the solid angle subtended by the disk \rev{and the filling factor of the disk emission \citep{Tripathi2017}}. The highly inclined disks studied here provide a fresh opportunity to test this hypothesis as inclination plays a key factor in setting this solid angle. Fig.\,\ref{fig:rdisk_lmm} shows that highly inclined disks are significantly fainter \rev{in both continuum and line emission} than disks seen at lower inclination for a given disk radius. Furthermore, within the latter sample, there is a clear gradient of disk flux with inclination \rev{for the continuum emission}, the lower inclination systems being systematically brighter. To further test this geometrical effect, we apply a simple correction to account for projection effects, i.e., we divide the measured flux by cos($i$). \rev{We note that this correction assumes an infinitely thin emitting layer, which may not be perfect for the CO emission or for the continuum emission when the inclination approaches 90\degr. Furthermore, small inclination uncertainties result in large errors on the correction factor. Nonetheless, this simple approach provides a purely geometrical correction that enables a comparison of the overall sample.} As the right panels of Fig.\,\ref{fig:rdisk_lmm} illustrate, the correlation between disk size and flux becomes much tighter and highly-inclined disks now reside along the same correlations as lower inclination systems. For highly-inclined disks, the confidence level in the correlation between millimeter flux and disks increases to 99.1\% \rev{and 99.5\%} ($r=0.66$ \rev{and $r=0.71$} from Pearson tests) for the continuum \rev{and line emission, respectively}. Overall, these results provide strong support for the optical depth interpretation of the radius-flux correlation \rev{for both continuum and line emission}.

\subsubsection{Vertical extents} 
\label{Sect:Discussion_vertical_extent}

The comparison of vertical extents between gas and dust populations is not as well documented as the radial extent. For the comparison between the scattered light dust and the gas, we added vertical size measurements of 3 Class II low inclination disks from \cite{Avenhaus_2018} and \citet{Pinte2018, Galloway-Sprietsma_2025}, respectively (Fig. \ref{fig:discuss_sizes}d). Contrary to the radial direction, a few disks are more extended vertically in scattered light than in gas, or close to $H_\mathrm{gas}$ $\approx$ $H_\mathrm{dust,\,\mu m}$. We stress again that both quantities represent the full vertical extension of the two tracers, not their scale height. We can attribute these results to less evolved disks with the majority of small grains in the upper layers of the disk. This is consistent with the finding that grains as large as 10\,$\mu$m are fully mixed to the gas up to the disk surface in Tau\,0420201 \citep{Duchene_2024}, for which $H_\mathrm{gas} \approx H_\mathrm{dust,\,\mu m}$ However, this is surprising since the scattering surface should shield the gas underneath from the stellar radiation, preventing photodissociation. On the other hand, the majority of disks are close to $H_\mathrm{gas}$ $\approx$ 2--3 $H_\mathrm{dust,\,\mu m}$. This suggests that the scattered light dust is not perfectly coupled to the gas and some $\mu$m-size grains of the order of a few $\mu$m may have started to settle toward the midplane. Alternatively, given its high optical depth, CO self-shielding could result in an extended molecular layer that reaches out above the scattered light surface. We did not notice a different behavior between low and high inclinations disks, although constraints on the vertical extent across a larger sample of disks are needed.

However, we can see that the millimeter dust is clearly settled in the midplane, indicating significant decoupling from the gas (see Fig. \ref{fig:discuss_sizes}e), much more pronounced than for the radial direction. The majority of the disks follow $H_\mathrm{gas}$ $\approx$ 3--6  $H_\mathrm{dust,mm}$, indicator of strong dust settling. This fact has already been reported \citep[e.g.][]{Villenave2020} when comparing thermal emission and scattered light. Instead, here, we present the first quantitative analysis by comparing directly gas and continuum on a significant sample of disks.

\begin{figure*}[h!]
        \centering        \includegraphics[width=0.44\textwidth]{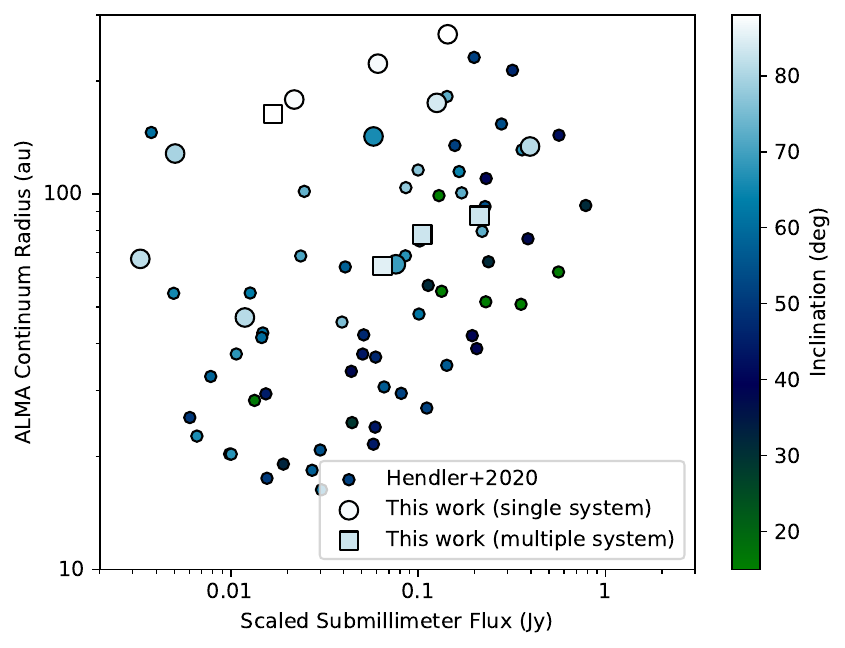} \includegraphics[width=0.44\textwidth]{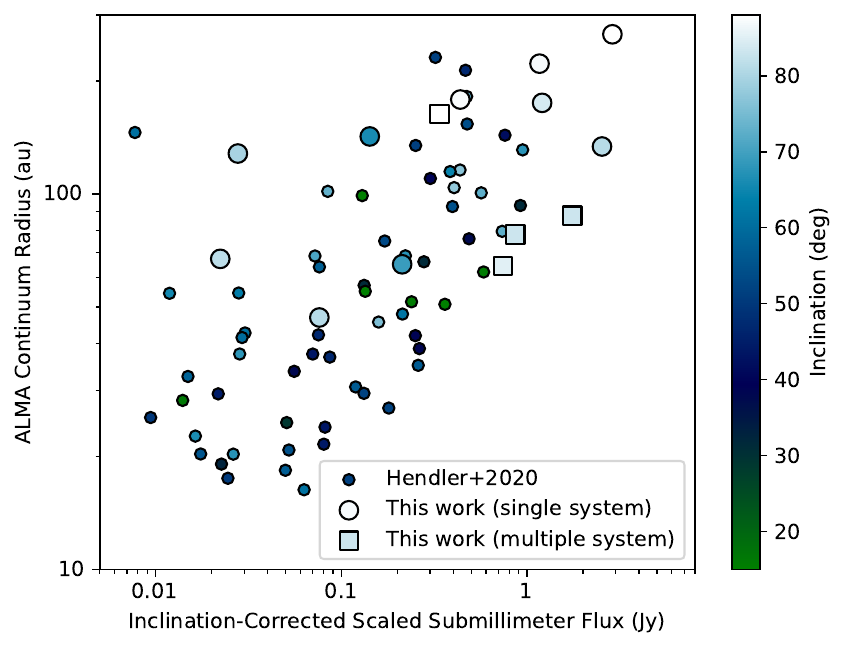}
        \includegraphics[width=0.44\textwidth]{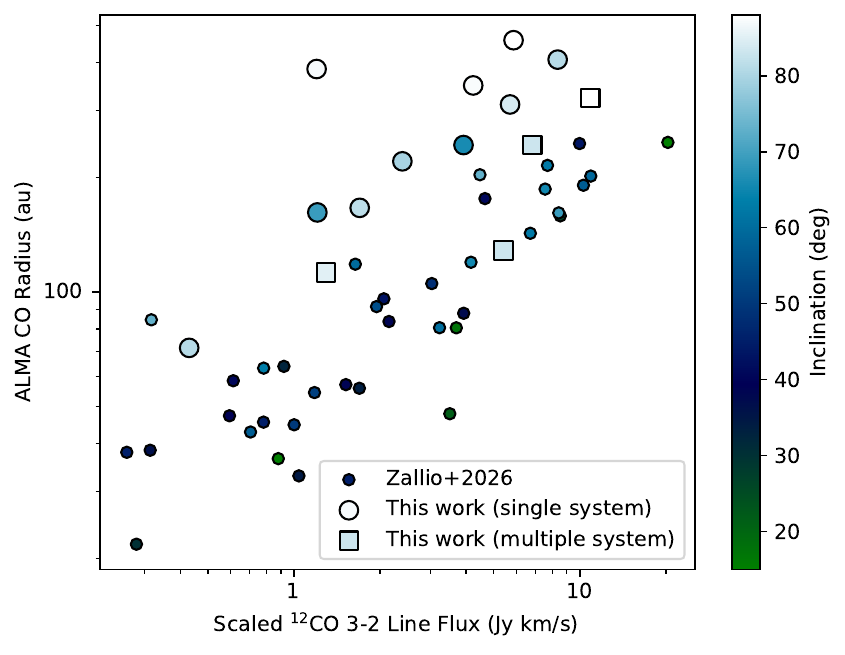} \includegraphics[width=0.44\textwidth]{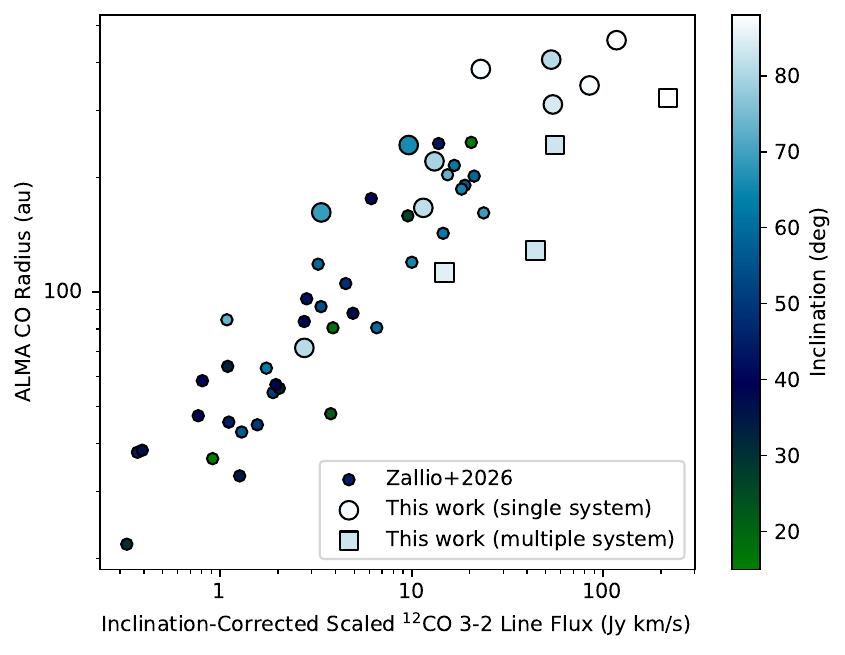}
        \caption{Disk radii measured in the ALMA continuum \rev{(top row) and $^{12}$CO 3--2 line (bottom row)} as a function of their submillimeter continuum \rev{and line flux} (scaled to 140\,pc for uniformity), respectively. Small circles represent low- to moderate-inclination disks from the literature \citep{Hendler_2020, Zallio_2026} while large circles and squares are highly inclined disks from this study in single and multiple systems, respectively.
        The disk inclination is color-coded. In the right column, the fluxes are further corrected for inclination, i.e., divided by cos($i$), to account for their projected shape.}
        \label{fig:rdisk_lmm}
\end{figure*}

\subsection{Implications of dynamical masses}

In this subsection, we discuss implications of dynamical masses on disks and stellar properties: gas and millimeter dust disk radius, aspect ratio ($h/r$) and spectral type. \rev{Despite significant uncertainties on the derived stellar masses, several notable trends emerge from our survey.}

\subsubsection{Disk properties} 
No correlation is found between the stellar mass and the disk size but one is found for the aspect ratio. The comparison between stellar mass and millimeter dust is shown in Fig. \ref{fig:size_vs_mass}. According to a Pearson test, we found no correlation between the stellar dynamical mass and the millimeter dust disk radius, with a $\approx$\,62$\%$ confidence level, which is too low to claim any correlation. This is in agreement with previous studies on less inclined disks \citep{Hendler_2020}. 
\final{\citet{Long2022} found a mild correlation between the gas disk radius and the stellar mass. However, we found that the gas size is even less correlated with the dynamical mass than the dust size. This could be a consequence of selection biases in either samples, but we also note that the correlation found by \cite{Long2022} is driven by the low-mass end of their sample (M < 0.25 M$_\odot$), which our sample does not cover.}

Fig. \ref{fig:mass_vs_hr} shows the stellar mass as a function of $h/r$ ratio defined here as the ratio between the vertical and radial gas sizes. As expected, we find an anti-correlation between the dynamical mass and the h/r ratio with disks around higher mass stars thinner than disks around lower mass stars. A Pearson test to check if the dynamical mass and the aspect ratio is correlated indicates that the two quantities
seem anti-correlated with a $\approx$ 91$\%$ confidence level and a correlation coefficient of $r=-0.49$. Indeed, higher mass stars have a stronger gravity pull, which tends to make the disk thinner. More massive stars are also more luminous, resulting in hotter dust, and thus pushing up the gas scale height, i.e., making the disk thicker. However, for pre-main-sequence stars, the effect of gravity is much steeper than that of luminosity, resulting in thinner disks around higher-mass stars \citep{Walker2004, Angelo2023}. We caution that the aspect ratio we discuss here is based on the full height of the gas-emitting layer, instead of its hydrostatic scale height. A thorough interpretation of this anti-correlation will require full radiative transfer models tailored to each system, as well as tomographic reconstruction of their vertical temperature profile \citep[e.g.,][]{Dutrey_2017, Flores_2021}, which is beyond the scope of this study. \rev{Finally, there is no significant correlation between the disk radius (dust and gas) and the aspect ratio. 
}

\begin{figure}[h!]
        \centering
         \includegraphics[width=0.46\textwidth]{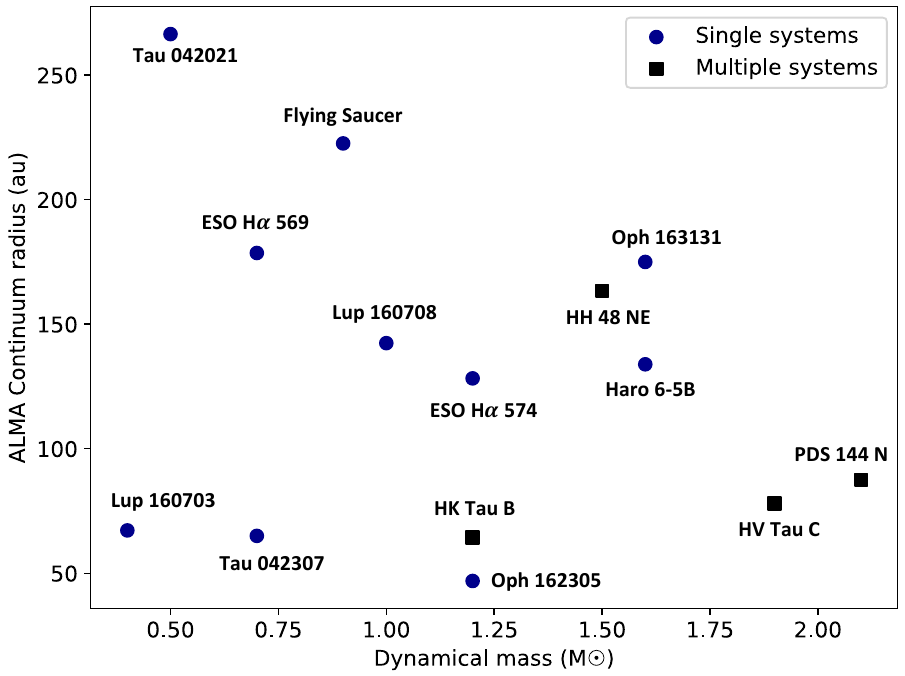}
        \caption{Millimeter dust disk radius as a function of stellar mass.}
        \label{fig:size_vs_mass}
\end{figure}

\subsubsection{Stellar properties} 
Fig. \ref{fig:mass_vs_spty} displays the dynamical mass as a function of the spectral type. We also add measurements from \citet{Simon_2019} derived with the DiskFit package method \citep{pietu_2007}. Our values are consistent with the reported trend of a decreasing dynamical mass for spectral type ranging from A2 to M4.75, although there is a large scatter in the literature sample, suggestive of significant uncertainties. However, a few disks constitute outliers: the K6 multiple system HV\,Tau\,C, the M1 edge-on system Tau\,042021, as well as the M4.75 grazing-angle system Tau\,042307. Outliers might be due to an incorrect estimate of the spectral type or an inaccurate estimation of the dynamical mass linked to the limitation of the method in the highly inclined configuration in some cases. 

By comparing our dynamical masses values with the ones in the literature, when available, we found our values systematically higher than those of the literature by 30--50\%, although our values are consistent within uncertainties, at the exception of Haro\,6-5B (see Table \ref{table:MASSES}). The dynamical mass estimated by \citet{Yokogawa_2002} for Haro\,6-5B is significantly lower (0.25 M$_\odot$ as opposed to our estimate of 1.6 M$_\odot$), and seems incompatible with both our PV-diagrams (Fig. \ref{fig:PV_DIAGRAMS_ALL}) and the K5 spectral type of the system. However, we note that the dynamical masses found for the edge-on disks in \citet{Simon_2019} might not be fully accurate because of the limitation of their method due to projection effects when the two CO surfaces are well-separated. Instead, this issue is irrelevant to our PV diagram analysis.

\begin{figure}[htb]
        \centering
        \includegraphics[width=0.45\textwidth]{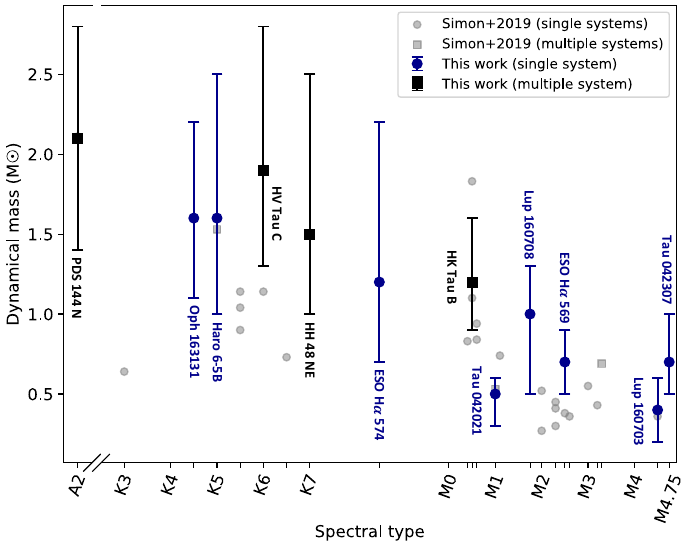}
        \caption{Dynamical mass as a function of spectral type. We averaged for objects which the spectral type is a range. The error bars represent conservative ($\approx3\sigma$) limits derived from our PV diagram analysis (see Sect. \ref{sect:dynamical_masses}).}
        \label{fig:mass_vs_spty}
\end{figure}

\begin{figure}[htb]
        \centering
        \includegraphics[width=0.45\textwidth]{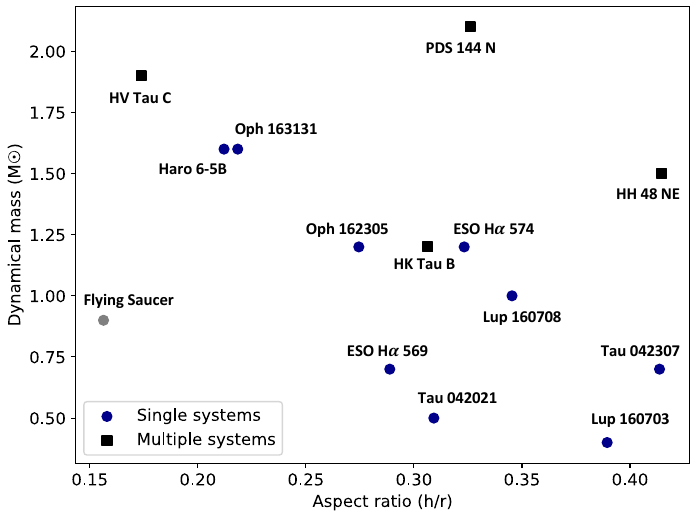}
        \caption{Dynamical mass as a function of aspect ratio ($h/r$), which represents the vertical extent over the radial extent of $^{12}$CO gas. Flying\,Saucer is shown in gray to represent the low SNR (see Sect. \ref{sect:results_alma}) and is not included in the statistical test presented in Sect. \ref{sect:dynamical_masses} .}
        \label{fig:mass_vs_hr}
\end{figure}

\subsection{Multiple systems}
\label{sect:multiple_systems}


The dust component of disks in multiple systems are expected to be smaller than disks in single systems due to dynamical interactions with the disks companions \citep{Manara_2019}, impacting planetary formation. However, we found no significant difference between single and binary stars within our sample in any tracer (see top panels of Fig. \ref{fig:size_vs_mass}). Moreover, as mentioned in Sect. \ref{sect:comparison_sizes}, HV\,Tau\,C and HH\,48\,NE also appear radially smaller in scattered light than in millimeter continuum which is surprising because their gas components are very extended. These small sizes could be explained as the result of stronger radial drift of $\mu$m-size grains than in single systems or to a shadowing effect. Hydrodynamical models indicate that gas disks should be truncated at $\approx$ 30$\%$ of the companion separation \citep{Artymowicz_1994, Zagaria_2021}. However, the binary separation is 2$\farcs$32 for HK\,Tau \citep{Rota_2022}, 5$\farcs$4 for PDS\,144 \citep{Perrin_2006}, 4$\farcs$04 for HV\,Tau \citep{Duchene_2010} and 2$\farcs$3 for HH\,48 \citep{Sturm2023_HH48_modeling}. This implies that the gas radius of the disk is below 30$\%$ of the separation for HK\,Tau and PDS\,144 but is higher for HV\,Tau and HH\,48. Thus, the small sizes of these disks can not be explained by tidal truncation alone. This has already been reported by \citet{Garufi2020}. One possible interpretation of the disks larger than their predicted truncation-driven size is that their real separation is much larger than the separation projected on the sky. Another possible explanation is more complex interactions. At each periastron passage, material can be stripped from the disk, leading to structures, such as tails, that extend out to much larger distances \citep{Cuello_2025}. We observe this in HH\,48 and it could also be happening in the HV\,Tau system. We also notice that HV\,Tau recently experienced a flyby by DO\,Tau \citep{Howard_2013}, which could also have stirred the HV\,Tau\,C disk.
Moreover, according to \citet{Rota_2022} there is no evidence of correlation between the size ratio and the projected separation, suggesting that there is no size dependence with the separation between components in the system. However, \citet{Rota_2022} found that the ratio of the dust to gas ratio in multiple stellar systems is lower compared to single stars, which is in agreement with what we found (see Fig. \ref{fig:discuss_sizes}a).


We also estimated the dynamical masses as well as their $^{12}$CO gas radial and vertical extents as we did for the primary stars for all disk-hosting companions: HK\,Tau\,A, PDS\,144\,S and HH\,48\,SW. 
All values are summarized in Appendix \ref{sect:BINARIES}. The primary star in a multiple system being the most massive component, we identify HK\,Tau\,B, HH\,48\,NE and PDS\,144\,N as the primaries in their respective systems, although it is more debatable for the PDS\,144 system because the two companions have almost the same masses (see App. B.5). Given the spectral types of the two components \citep[A2 and A5][]{Vieira2003}, we considered PDS\,144\,N as the primary. Primary stars generally host brighter disks than their companions \citep{Akeson_2019, Manara_2019}, which is consistent with our findings — except for HK\,Tau, for which both the continuum and line flux are brighter around the secondary. We also found that the millimeter dust radial extent is larger in the primary than in secondary component, which is in agreement with \citet{Manara_2019}, as is also the case for the $^{12}$CO gas radial extent. However, this is the opposite for the $^{12}$CO gas vertical extent which is more extended in the secondary component than in the primary component, which could be due to a larger scale height due to the lower stellar mass of the secondaries. Even though the disks around companions are also highly inclined, there is no clear tendency for coplanar disks in the systems studied here, which is expected given their large separations \citep{Jensen2014, Rota_2022}.

We also estimated an inclination of the disks around companions based on their millimeter continuum radial and vertical extents. Their inclinations also suggest highly inclined disks ($\gtrsim75\degr$), even though the star is directly visible in the optical/NIR, which suggest lower inclinations. Only the HK\,Tau\,A disk has a moderate inclination in the mm. The fact that some disks could appear highly inclined at millimeter wavelengths and less inclined in scattered-light might be explained \rev{by a combination of a smaller scale height and a grazing-angle configuration with an inclination a few degrees too low to fully obscure the star.} 
This can be confirmed in these systems with high-contrast scattered light imaging of the disks around these companions. \rev{Indeed, \cite{garufi2026} recently presented the first scattered light image with ground-based adaptive optics of PDS\,144\,S, which is well aligned with the ALMA continuum and CO maps, confirming its highly inclined nature.}


\subsection{Grazing-angle systems}

The grazing-angle disks constitute a category of highly inclined disks that is different from edge-on disks. Four disks in our sample present a grazing-angle geometry: Tau\,042307, Haro\,6-5B, Lup\,160703 and Lup\,160708. Their millimeter continuum and $^{12}$CO gas images are not as easy recognizable as in optical/NIR image. However, as mentioned in Sect. \ref{sect:results_alma} some distinguishable features can be present in the grazing-angle disks images allowing to disentangle between the edge-on disks images.

These disks represent the best compromise between pole-on disks and edge-on disks for which the radial and vertical structures, respectively, are accessible, allowing to study both the radial and the vertical structure of the continuum emission of disks. The grazing-angle geometry also provides an ideal configuration to probe the vertical temperature gradient through the CO-emitting layer, as for less inclined disks \citep[e.g.,][]{Pinte2018}. Indeed, the lower nebula appears fainter than the upper nebula in optical/NIR due to the combination of the high optical depth of the $^{12}$CO line and the large vertical extent of, and temperature gradient through, the CO-emitting layer.
Therefore, since methods developed for edge-on disks are also applicable to these systems, grazing-angle disks can be used as benchmarks to bridge the gap between pole-on and edge-on disks, ultimately providing a global context.

Grazing-angle disks are also of great interest in the study of ices because this geometry is the only one for which we can observe both scattering and absorption components \citep{Martinien_2024, Martinien_2025b, Martinien_2025a}, allowing an estimation of ices abundances. Thus, they offer a unique opportunity -- through modeling -- to compare the solid state,
with James Webb Space Telescope spectroscopic observations, and gaseous CO abundances, with ALMA observations. This can address whether the two can be co-located, i.e., partial sublimation. Unfortunately, high resolution, high SNR data are missing for most  grazing angle for now.

\section{Conclusion}
\label{sect:conclusions}

This study presents a high-resolution gas and continuum survey of highly inclined protoplanetary disks with ALMA. From these observations, we perform a comparison of disks sizes in millimeter continuum, $^{12}$CO gas and optical/NIR observations along the radial and vertical direction, within a sample of highly inclined disks. It also provides the first dynamical mass estimates for most of the objects using the position–velocity diagram method. Our conclusions are as follows:

\begin{enumerate}
    \item Most of the radial disks sizes in our sample follow: $R_\mathrm{gas}$ > $R_\mathrm{dust,\,\mu m}$ > $R_\mathrm{dust, \, mm}$, with the exception of 3 disks (including 2 in multiples systems) for which $R_\mathrm{dust, \, mm}$ is greater than $R_\mathrm{dust,\,\mu m}$. Highly inclined disks follow the same trend as less inclined disks, albeit with a lower value of $R_\mathrm{gas}$ / $R_\mathrm{dust, \, mm}$ $\approx$ 1.5--2. While this is broadly consistent with optical depth effects in disks with a uniform gas and dust mixture, we 
    stress that radial drift could nonetheless be active, especially in systems with $R_\mathrm{dust, \, mm} < R_\mathrm{dust,\,\mu m}$.
    Highly inclined disks provide a more robust estimate of the outer disk radius because of line-of-sight integration. 
    \item Along the vertical direction, all disks in our sample follow $H_\mathrm{gas}$ >> $H_\mathrm{dust, mm}$. Most of disks follow $H_\mathrm{gas}$ >  $H_\mathrm{dust,\,\mu m}$, but a few disks follow $H_\mathrm{gas}$ $\approxeq$ $H_\mathrm{dust,\,\mu m}$ or even $H_\mathrm{gas}$ < $H_\mathrm{dust,\,\mu m}$. This suggests that the millimeter dust is vertically settled toward the midplane, while the $\mu$m-sized dust is not fully coupled to the gas. 
    \item We revisited the long-known correlation between the size and millimeter continuum \rev{and line} emission of disks. Highly-inclined disks are systematically fainter than disks at lower inclinations, in a way that is consistent with the smaller solid angle they subtend. This reinforces the idea that the correlation is primarily a consequence of the fact that \rev{both} the continuum \rev{and line} emission are optically thick in most disks.
    \item We found an anti-correlation between the dynamical mass and the $h/r$ ratio, consistent with the dominating effect of gravity in setting the height of a disk, but no correlation between the dynamical mass and the disk radius.
    \item We also studied the companions in multiple systems. Even though they are directly visible in optical images, 2 of the 3 companions also host highly inclined disks ($\gtrsim75\degr$), which could be due to a tilt or a warp, easily induced in multiple systems. The disk associated with the most massive component, always the edge-on disk component, are more radially extended in all tracers. Conversely, disks around the lower-mass components are more vertically extended in their gas component but similarly thick in their continuum emission.
    \item We can distinguish between edge-on and grazing-angle disks, which can offer a unique viewing geometry, providing the best compromise to study radial and vertical structure of disks.

\end{enumerate}

Higher resolution observations are needed to better constrain the dynamical masses and a larger sample is needed to strengthen comparison between grazing-angle, edge-on and multiple systems among a sample of highly inclined disks. In this context, the ongoing DiskStrat ALMA Large Program (Project 2024.1.01212.L, PI: R. Le Gal) will provide critical observations
to improve our characterization and understanding of highly inclined disks.

\begin{acknowledgements}
This project has received funding from the European Research Council (ERC) under the European Union’s Horizon Europe research and innovation program (grant agreement No. 101053020, project Dust2Planets, PI: F. M\'enard). AR has received funding from the Royal Society through a University Research Fellowship grant number URF\textbackslash R1\textbackslash 241791 and from the UK Science and Technology research Council (STFC) via the consolidated grant ST/W000997/1.
\end{acknowledgements}

\bibliography{BibLau.bib}

@ARTICLE{Akeson_2019,
       author = {{Akeson}, Rachel L. and {Jensen}, Eric L.~N. and {Carpenter}, John and {Ricci}, Luca and {Laos}, Emily and {Nogueira}, Natasha F. and {Suen-Lewis}, Emma M.},
        title = "{Resolved Young Binary Systems and Their Disks}",
      journal = {\apj},
         year = 2019,
        month = feb,
       volume = {872},
       number = {2},
          eid = {158},
        pages = {158},
          doi = {10.3847/1538-4357/aaff6a},
archivePrefix = {arXiv},
       eprint = {1901.05029},
 primaryClass = {astro-ph.SR},
       adsurl = {https://ui.adsabs.harvard.edu/abs/2019ApJ...872..158A}
}

@ARTICLE{Andrews_2018,
       author = {{Andrews}, Sean M. and {Terrell}, Marie and {Tripathi}, Anjali and {Ansdell}, Megan and {Williams}, Jonathan P. and {Wilner}, David J.},
        title = "{Scaling Relations Associated with Millimeter Continuum Sizes in Protoplanetary Disks}",
      journal = {\apj},
         year = 2018,
        month = oct,
       volume = {865},
       number = {2},
          eid = {157},
        pages = {157},
          doi = {10.3847/1538-4357/aadd9f},
archivePrefix = {arXiv},
       eprint = {1808.10510},
 primaryClass = {astro-ph.EP},
       adsurl = {https://ui.adsabs.harvard.edu/abs/2018ApJ...865..157A}
}

@ARTICLE{Andrews2020,
       author = {{Andrews}, Sean M.},
        title = "{Observations of Protoplanetary Disk Structures}",
      journal = {\araa},
         year = 2020,
        month = aug,
       volume = {58},
        pages = {483-528},
          doi = {10.1146/annurev-astro-031220-010302},
archivePrefix = {arXiv},
       eprint = {2001.05007},
 primaryClass = {astro-ph.EP},
       adsurl = {https://ui.adsabs.harvard.edu/abs/2020ARA&A..58..483A}
}

@article{Angelo2023,
 adsurl = {https://ui.adsabs.harvard.edu/abs/2023ApJ...945..130A},
 archiveprefix = {arXiv},
 author = {{Angelo}, Isabel and {Duchene}, Gaspard and {Stapelfeldt}, Karl and {Telkamp}, Zoie and {M{\'e}nard}, Fran{\c{c}}ois and {Padgett}, Deborah and {Van der Plas}, Gerrit and {Villenave}, Marion and {Pinte}, Christophe and {Wolff}, Schuyler and {Fischer}, William J. and {Perrin}, Marshall D.},
 doi = {10.3847/1538-4357/acbb01},
 eid = {130},
 eprint = {2302.04891},
 journal = {\apj},
 month = {March},
 number = {2},
 pages = {130},
 primaryclass = {astro-ph.EP},
 title = {{Demographics of Protoplanetary Disks: A Simulated Population of Edge-on Systems}},
 volume = {945},
 year = {2023}
}

@ARTICLE{Ansdell_2018,
       author = {{Ansdell}, M. and {Williams}, J.~P. and {Trapman}, L. and {van Terwisga}, S.~E. and {Facchini}, S. and {Manara}, C.~F. and {van der Marel}, N. and {Miotello}, A. and {Tazzari}, M. and {Hogerheijde}, M. and {Guidi}, G. and {Testi}, L. and {van Dishoeck}, E.~F.},
        title = "{ALMA Survey of Lupus Protoplanetary Disks. II. Gas Disk Radii}",
      journal = {\apj},
         year = 2018,
        month = may,
       volume = {859},
       number = {1},
          eid = {21},
        pages = {21},
          doi = {10.3847/1538-4357/aab890},
archivePrefix = {arXiv},
       eprint = {1803.05923},
 primaryClass = {astro-ph.EP},
       adsurl = {https://ui.adsabs.harvard.edu/abs/2018ApJ...859...21A}
}

@ARTICLE{Artymowicz_1994,
       author = {{Artymowicz}, Pawel and {Lubow}, Stephen H.},
        title = "{Dynamics of Binary-Disk Interaction. I. Resonances and Disk Gap Sizes}",
      journal = {\apj},
         year = 1994,
        month = feb,
       volume = {421},
        pages = {651},
          doi = {10.1086/173679},
       adsurl = {https://ui.adsabs.harvard.edu/abs/1994ApJ...421..651A}
}

@ARTICLE{Avenhaus_2018,
       author = {{Avenhaus}, Henning and {Quanz}, Sascha P. and {Garufi}, Antonio and {Perez}, Sebastian and {Casassus}, Simon and {Pinte}, Christophe and {Bertrang}, Gesa H. -M. and {Caceres}, Claudio and {Benisty}, Myriam and {Dominik}, Carsten},
        title = "{Disks around T Tauri Stars with SPHERE (DARTTS-S). I. SPHERE/IRDIS Polarimetric Imaging of Eight Prominent T Tauri Disks}",
      journal = {\apj},
         year = 2018,
        month = aug,
       volume = {863},
       number = {1},
          eid = {44},
        pages = {44},
          doi = {10.3847/1538-4357/aab846},
archivePrefix = {arXiv},
       eprint = {1803.10882},
 primaryClass = {astro-ph.SR},
       adsurl = {https://ui.adsabs.harvard.edu/abs/2018ApJ...863...44A}
}

@ARTICLE{Barriere_Fouchet_2005,
       author = {{Barri{\`e}re-Fouchet}, L. and {Gonzalez}, J.-F. and {Murray}, J.~R. and {Humble}, R.~J. and {Maddison}, S.~T.},
        title = "{Dust distribution in protoplanetary disks. Vertical settling and radial migration}",
      journal = {\aap},
         year = 2005,
        month = nov,
       volume = {443},
       number = {1},
        pages = {185-194},
          doi = {10.1051/0004-6361:20042249},
archivePrefix = {arXiv},
       eprint = {astro-ph/0508452},
 primaryClass = {astro-ph},
       adsurl = {https://ui.adsabs.harvard.edu/abs/2005A&A...443..185B}
}

@INPROCEEDINGS{benisty2022optical,
       author = {{Benisty}, M. and {Dominik}, C. and {Follette}, K. and {Garufi}, A. and {Ginski}, C. and {Hashimoto}, J. and {Keppler}, M. and {Kley}, W. and {Monnier}, J.},
        title = "{Optical and Near-infrared View of Planet-forming Disks and Protoplanets}",
    booktitle = {Protostars and Planets VII},
         year = 2023,
       editor = {{Inutsuka}, S. and {Aikawa}, Y. and {Muto}, T. and {Tomida}, K. and {Tamura}, M.},
       series = {Astronomical Society of the Pacific Conference Series},
       volume = {534},
        month = jul,
        pages = {605},
          doi = {10.48550/arXiv.2203.09991},
archivePrefix = {arXiv},
       eprint = {2203.09991},
 primaryClass = {astro-ph.EP},
       adsurl = {https://ui.adsabs.harvard.edu/abs/2023ASPC..534..605B}
}

@ARTICLE{Birnstiel2014,
       author = {{Birnstiel}, Tilman and {Andrews}, Sean M.},
        title = "{On the Outer Edges of Protoplanetary Dust Disks}",
      journal = {\apj},
         year = 2014,
        month = jan,
       volume = {780},
       number = {2},
          eid = {153},
        pages = {153},
          doi = {10.1088/0004-637X/780/2/153},
archivePrefix = {arXiv},
       eprint = {1311.5222},
 primaryClass = {astro-ph.EP},
       adsurl = {https://ui.adsabs.harvard.edu/abs/2014ApJ...780..153B}
}

@ARTICLE{Cox_2017,
       author = {{Cox}, Erin G. and {Harris}, Robert J. and {Looney}, Leslie W. and {Chiang}, Hsin-Fang and {Chandler}, Claire and {Kratter}, Kaitlin and {Li}, Zhi-Yun and {Perez}, Laura and {Tobin}, John J.},
        title = "{Protoplanetary Disks in {\ensuremath{\rho}} Ophiuchus as Seen from ALMA}",
      journal = {\apj},
         year = 2017,
        month = dec,
       volume = {851},
       number = {2},
          eid = {83},
        pages = {83},
          doi = {10.3847/1538-4357/aa97e2},
archivePrefix = {arXiv},
       eprint = {1711.03974},
 primaryClass = {astro-ph.SR},
       adsurl = {https://ui.adsabs.harvard.edu/abs/2017ApJ...851...83C}
}

@ARTICLE{Cuello_2025,
       author = {{Cuello}, Nicol{\'a}s and {Alaguero}, Antoine and {Poblete}, Pedro P.},
        title = "{Circumstellar and Circumbinary Discs in Multiple Stellar Systems}",
      journal = {Symmetry},
         year = 2025,
        month = feb,
       volume = {17},
       number = {3},
          eid = {344},
        pages = {344},
          doi = {10.3390/sym17030344},
archivePrefix = {arXiv},
       eprint = {2501.19249},
 primaryClass = {astro-ph.EP},
       adsurl = {https://ui.adsabs.harvard.edu/abs/2025Symm...17..344C}
}

@ARTICLE{Czekala2016,
       author = {{Czekala}, I. and {Andrews}, S.~M. and {Torres}, G. and {Jensen}, E.~L.~N. and {Stassun}, K.~G. and {Wilner}, D.~J. and {Latham}, D.~W.},
        title = "{A Disk-based Dynamical Constraint on the Mass of the Young Binary DQ Tau}",
      journal = {\apj},
         year = 2016,
        month = feb,
       volume = {818},
       number = {2},
          eid = {156},
        pages = {156},
          doi = {10.3847/0004-637X/818/2/156},
archivePrefix = {arXiv},
       eprint = {1601.03806},
 primaryClass = {astro-ph.SR},
       adsurl = {https://ui.adsabs.harvard.edu/abs/2016ApJ...818..156C}
}

@article{Dartois2024,
 adsurl = {https://ui.adsabs.harvard.edu/abs/2024NatAs.tmp....6D},
 author = {{Dartois}, E. and {Noble}, J.~A. and {Caselli}, P. and {Fraser}, H.~J. and {Jim{\'e}nez-Serra}, I. and {Mat{\'e}}, B. and {McClure}, M.~K. and {Melnick}, G.~J. and {Pendleton}, Y.~J. and {Shimonishi}, T. and {Smith}, Z.~L. and {Sturm}, J.~A. and {Taillard}, A. and {Wakelam}, V. and {Boogert}, A.~C.~A. and {Drozdovskaya}, M.~N. and {Erkal}, J. and {Harsono}, D. and {Herrero}, V.~J. and {Ioppolo}, S. and {Linnartz}, H. and {McGuire}, B.~A. and {Perotti}, G. and {Qasim}, D. and {Rocha}, W.~R.~M.},
 doi = {10.1038/s41550-023-02155-x},
 journal = {Nature Astronomy},
 month = {January},
 title = {{Spectroscopic sizing of interstellar icy grains with JWST}},
 year = {2024}
}

@ARTICLE{Deng_2025,
       author = {{Deng}, Dingshan and {Vioque}, Miguel and {Pascucci}, Ilaria and {P{\'e}rez}, Laura M. and {Zhang}, Ke and {Kurtovic}, Nicol{\'a}s and {Trapman}, Leon and {TorresVillanueva}, Estephani E. and {Agurto-Gangas}, Carolina and {Carpenter}, John and {Pinilla}, Paola and {Gorti}, Uma and {Tabone}, Beno{\^\i}t and {Sierra}, Anibal and {Rosotti}, Giovanni P. and {Cieza}, Lucas A. and {Anania}, Rossella and {Gonz{\'a}lez-Ruilova}, Camilo and {Hogerheijde}, Michiel R. and {Miley}, James and {Ruiz-Rodriguez}, Dary A. and {Ruaud}, Maxime and {Schwarz}, Kamber},
        title = "{The ALMA Survey of Gas Evolution of PROtoplanetary Disks (AGE-PRO). III. Dust and Gas Disk Properties in the Lupus Star-forming Region}",
      journal = {\apj},
         year = 2025,
        month = aug,
       volume = {989},
       number = {1},
          eid = {3},
        pages = {3},
          doi = {10.3847/1538-4357/add43a},
archivePrefix = {arXiv},
       eprint = {2506.10734},
 primaryClass = {astro-ph.EP},
       adsurl = {https://ui.adsabs.harvard.edu/abs/2025ApJ...989....3D}
}

@ARTICLE{Duchene_2010,
       author = {{Duch{\^e}ne}, G. and {McCabe}, C. and {Pinte}, C. and {Stapelfeldt}, K.~R. and {M{\'e}nard}, F. and {Duvert}, G. and {Ghez}, A.~M. and {Maness}, H.~L. and {Bouy}, H. and {Barrado y Navascu{\'e}s}, D. and {Morales-Calder{\'o}n}, M. and {Wolf}, S. and {Padgett}, D.~L. and {Brooke}, T.~Y. and {Noriega-Crespo}, A.},
        title = "{Panchromatic Observations and Modeling of the HV Tau C Edge-on Disk}",
      journal = {\apj},
         year = 2010,
        month = mar,
       volume = {712},
       number = {1},
        pages = {112-129},
          doi = {10.1088/0004-637X/712/1/112},
archivePrefix = {arXiv},
       eprint = {0911.3445},
 primaryClass = {astro-ph.SR},
       adsurl = {https://ui.adsabs.harvard.edu/abs/2010ApJ...712..112D}
}

@ARTICLE{Duchene_2024,
       author = {{Duch{\^e}ne}, Gaspard and {M{\'e}nard}, Fran{\c{c}}ois and {Stapelfeldt}, Karl R. and {Villenave}, Marion and {Wolff}, Schuyler G. and {Perrin}, Marshall D. and {Pinte}, Christophe and {Tazaki}, Ryo and {Padgett}, Deborah L.},
        title = "{JWST Imaging of Edge-on Protoplanetary Disks. I. Fully Vertically Mixed 10 {\ensuremath{\mu}}m Grains in the Outer Regions of a 1000 au Disk}",
      journal = {\aj},
         year = 2024,
        month = feb,
       volume = {167},
       number = {2},
          eid = {77},
        pages = {77},
          doi = {10.3847/1538-3881/acf9a7},
archivePrefix = {arXiv},
       eprint = {2309.07040},
 primaryClass = {astro-ph.EP},
       adsurl = {https://ui.adsabs.harvard.edu/abs/2024AJ....167...77D}
}

@ARTICLE{Dutrey_2017,
       author = {{Dutrey}, A. and {Guilloteau}, S. and {Pi{\'e}tu}, V. and {Chapillon}, E. and {Wakelam}, V. and {Di Folco}, E. and {Stoecklin}, T. and {Denis-Alpizar}, O. and {Gorti}, U. and {Teague}, R. and {Henning}, T. and {Semenov}, D. and {Grosso}, N.},
        title = "{The Flying Saucer: Tomography of the thermal and density gas structure of an edge-on protoplanetary disk}",
      journal = {\aap},
         year = 2017,
        month = nov,
       volume = {607},
          eid = {A130},
        pages = {A130},
          doi = {10.1051/0004-6361/201730645},
archivePrefix = {arXiv},
       eprint = {1706.02608},
 primaryClass = {astro-ph.SR},
       adsurl = {https://ui.adsabs.harvard.edu/abs/2017A&A...607A.130D}
}

@ARTICLE{Dutrey_2025,
       author = {{Dutrey}, A. and {Denis-Alpizar}, O. and {Guilloteau}, S. and {Foucher}, C. and {Gavino}, S. and {Semenov}, D. and {Pietu}, V. and {Chapillon}, E. and {Testi}, L. and {Dartois}, E. and {DiFolco}, E. and {Furuya}, K. and {Gorti}, U. and {Grosso}, N. and {Henning}, Th. and {Hur{\'e}}, J.~M. and {K{\'o}sp{\'a}l}, {\'A}. and {Le Petit}, F. and {Majumdar}, L. and {Meshaka}, R. and {Nomura}, H. and {Phuong}, N.~T. and {Ruaud}, M. and {Tang}, Y.~W. and {Wolf}, S.},
        title = "{Edge-On Disk Study (EODS) III: Molecular Stratification in the Flying Saucer Disk}",
      journal = {arXiv e-prints},
         year = 2025,
        month = sep,
          eid = {arXiv:2509.26033},
        pages = {arXiv:2509.26033},
          doi = {10.48550/arXiv.2509.26033},
archivePrefix = {arXiv},
       eprint = {2509.26033},
 primaryClass = {astro-ph.EP},
       adsurl = {https://ui.adsabs.harvard.edu/abs/2025arXiv250926033D}
}

@ARTICLE{Facchini_2017,
       author = {{Facchini}, S. and {Birnstiel}, T. and {Bruderer}, S. and {van Dishoeck}, E.~F.},
        title = "{Different dust and gas radial extents in protoplanetary disks: consistent models of grain growth and CO emission}",
      journal = {\aap},
         year = 2017,
        month = sep,
       volume = {605},
          eid = {A16},
        pages = {A16},
          doi = {10.1051/0004-6361/201630329},
archivePrefix = {arXiv},
       eprint = {1705.06235},
 primaryClass = {astro-ph.SR},
       adsurl = {https://ui.adsabs.harvard.edu/abs/2017A&A...605A..16F}
}

@ARTICLE{Flores_2021,
       author = {{Flores}, C. and {Duch{\^e}ne}, G. and {Wolff}, S. and {Villenave}, M. and {Stapelfeldt}, K. and {Williams}, J.~P. and {Pinte}, C. and {Padgett}, D. and {Connelley}, M.~S. and {van der Plas}, G. and {M{\'e}nard}, F. and {Perrin}, M.~D.},
        title = "{The Anatomy of an Unusual Edge-on Protoplanetary Disk. II. Gas Temperature and a Warm Outer Region}",
      journal = {\aj},
         year = 2021,
        month = may,
       volume = {161},
       number = {5},
          eid = {239},
        pages = {239},
          doi = {10.3847/1538-3881/abeb1e},
archivePrefix = {arXiv},
       eprint = {2103.02666},
 primaryClass = {astro-ph.SR},
       adsurl = {https://ui.adsabs.harvard.edu/abs/2021AJ....161..239F}
}

@ARTICLE{Galloway-Sprietsma_2025,
       author = {{Galloway-Sprietsma}, Maria and {Bae}, Jaehan and {Izquierdo}, Andr{\'e}s F. and {Stadler}, Jochen and {Longarini}, Cristiano and {Teague}, Richard and {Andrews}, Sean M. and {Winter}, Andrew J. and {Benisty}, Myriam and {Facchini}, Stefano and {Rosotti}, Giovanni and {Zawadzki}, Brianna and {Pinte}, Christophe and {Fasano}, Daniele and {Barraza-Alfaro}, Marcelo and {Cataldi}, Gianni and {Cuello}, Nicol{\'a}s and {Curone}, Pietro and {Czekala}, Ian and {Flock}, Mario and {Fukagawa}, Misato and {Gardner}, Charles H. and {Garg}, Himanshi and {Hall}, Cassandra and {Huang}, Jane and {Ilee}, John D. and {Kanagawa}, Kazuhiro and {Lesur}, Geoffroy and {Lodato}, Giuseppe and {Loomis}, Ryan A. and {Menard}, Francois and {Orihara}, Ryuta and {Price}, Daniel J. and {Wafflard-Fernandez}, Gaylor and {Wilner}, David J. and {W{\"o}lfer}, Lisa and {Yen}, Hsi-Wei and {Yoshida}, Tomohiro C.},
        title = "{exoALMA. V. Gaseous Emission Surfaces and Temperature Structures}",
      journal = {\apjl},
         year = 2025,
        month = may,
       volume = {984},
       number = {1},
          eid = {L10},
        pages = {L10},
          doi = {10.3847/2041-8213/adc437},
archivePrefix = {arXiv},
       eprint = {2504.19902},
 primaryClass = {astro-ph.EP},
       adsurl = {https://ui.adsabs.harvard.edu/abs/2025ApJ...984L..10G}
}

@article{Garufi2020,
 adsurl = {https://ui.adsabs.harvard.edu/abs/2020A&A...633A..82G},
 archiveprefix = {arXiv},
 author = {{Garufi}, A. and {Avenhaus}, H. and {P{\'e}rez}, S. and {Quanz}, S.~P. and {van Holstein}, R.~G. and {Bertrang}, G.~H. -M. and {Casassus}, S. and {Cieza}, L. and {Principe}, D.~A. and {van der Plas}, G. and {Zurlo}, A.},
 doi = {10.1051/0004-6361/201936946},
 eid = {A82},
 eprint = {1911.10853},
 journal = {\aap},
 month = {January},
 pages = {A82},
 primaryclass = {astro-ph.EP},
 title = {{Disks Around T Tauri Stars with SPHERE (DARTTS-S). II. Twenty-one new polarimetric images of young stellar disks}},
 volume = {633},
 year = {2020}
}

@ARTICLE{Garufi2026,
       author = {{Garufi}, Antonio and {Ginski}, Christian and {Benisty}, Myriam and {Vioque}, Miguel and {Winter}, Andrew and {Huang}, Jane and {Manara}, Carlo Felice and {Dominik}, Carsten},
        title = "{Planet-forming disks and their environment across regions and time from the full NIR census}",
      journal = {arXiv e-prints},
         year = 2026,
        month = mar,
          eid = {arXiv:2603.01703},
        pages = {arXiv:2603.01703},
          doi = {10.48550/arXiv.2603.01703},
archivePrefix = {arXiv},
       eprint = {2603.01703},
 primaryClass = {astro-ph.SR},
       adsurl = {https://ui.adsabs.harvard.edu/abs/2026arXiv260301703G}
}

@ARTICLE{Hendler_2020,
       author = {{Hendler}, Nathanial and {Pascucci}, Ilaria and {Pinilla}, Paola and {Tazzari}, Marco and {Carpenter}, John and {Malhotra}, Renu and {Testi}, Leonardo},
        title = "{The Evolution of Dust Disk Sizes from a Homogeneous Analysis of 1-10 Myr old Stars}",
      journal = {\apj},
         year = 2020,
        month = jun,
       volume = {895},
       number = {2},
          eid = {126},
        pages = {126},
          doi = {10.3847/1538-4357/ab70ba},
archivePrefix = {arXiv},
       eprint = {2001.02666},
 primaryClass = {astro-ph.EP},
       adsurl = {https://ui.adsabs.harvard.edu/abs/2020ApJ...895..126H}
}

@ARTICLE{Hernandez_2008,
       author = {{Hern{\'a}ndez}, Jes{\'u}s and {Hartmann}, Lee and {Calvet}, Nuria and {Jeffries}, R.~D. and {Gutermuth}, R. and {Muzerolle}, J. and {Stauffer}, J.},
        title = "{A Spitzer View of Protoplanetary Disks in the {\ensuremath{\gamma}} Velorum Cluster}",
      journal = {\apj},
         year = 2008,
        month = oct,
       volume = {686},
       number = {2},
        pages = {1195-1208},
          doi = {10.1086/591224},
archivePrefix = {arXiv},
       eprint = {0806.2639},
 primaryClass = {astro-ph},
       adsurl = {https://ui.adsabs.harvard.edu/abs/2008ApJ...686.1195H}}

@ARTICLE{Hornbeck_2012,
       author = {{Hornbeck}, J.~B. and {Grady}, C.~A. and {Perrin}, M.~D. and {Wisniewski}, J.~P. and {Tofflemire}, B.~M. and {Brown}, A. and {Holtzman}, J.~A. and {Arraki}, K. and {Hamaguchi}, K. and {Woodgate}, B. and {Petre}, R. and {Daly}, B. and {Grogin}, N.~A. and {Bonfield}, D.~G. and {Williger}, G.~M. and {Lauroesch}, J.~T.},
        title = "{PDS 144: The First Confirmed Herbig Ae-Herbig Ae Wide Binary}",
      journal = {\apj},
         year = 2012,
        month = jan,
       volume = {744},
       number = {1},
          eid = {54},
        pages = {54},
          doi = {10.1088/0004-637X/744/1/54},
archivePrefix = {arXiv},
       eprint = {1201.0173},
 primaryClass = {astro-ph.SR},
       adsurl = {https://ui.adsabs.harvard.edu/abs/2012ApJ...744...54H}
}

@ARTICLE{Howard_2013,
       author = {{Howard}, Christian D. and {Sandell}, G{\"o}ran and {Vacca}, William D. and {Duch{\^e}ne}, Gaspard and {Mathews}, Geoffrey and {Augereau}, Jean-Charles and {Barrado}, David and {Dent}, William R.~F. and {Eiroa}, Carlos and {Grady}, Carol and {Kamp}, Inga and {Meeus}, Gwendolyn and {M{\'e}nard}, Francois and {Pinte}, Christophe and {Podio}, Linda and {Riviere-Marichalar}, Pablo and {Roberge}, Aki and {Thi}, Wing-Fai and {Vicente}, Silvia and {Williams}, Jonathan P.},
        title = "{Herschel/PACS Survey of Protoplanetary Disks in Taurus/Auriga{\textemdash}Observations of [O I] and [C II], and Far-infrared Continuum}",
      journal = {\apj},
         year = 2013,
        month = oct,
       volume = {776},
       number = {1},
          eid = {21},
        pages = {21},
          doi = {10.1088/0004-637X/776/1/21},
archivePrefix = {arXiv},
       eprint = {1308.6019},
 primaryClass = {astro-ph.GA},
       adsurl = {https://ui.adsabs.harvard.edu/abs/2013ApJ...776...21H}
}

@ARTICLE{jensen2014,
       author = {{Jensen}, Eric L.~N. and {Akeson}, Rachel},
        title = "{Misaligned protoplanetary disks in a young binary star system}",
      journal = {\nat},
         year = 2014,
        month = jul,
       volume = {511},
       number = {7511},
        pages = {567-569},
          doi = {10.1038/nature13521},
archivePrefix = {arXiv},
       eprint = {1407.8211},
 primaryClass = {astro-ph.SR},
       adsurl = {https://ui.adsabs.harvard.edu/abs/2014Natur.511..567J}
}

@article{Johansen2007,
 adsurl = {https://ui.adsabs.harvard.edu/abs/2007ApJ...662..627J},
 archiveprefix = {arXiv},
 author = {{Johansen}, A. and {Youdin}, A.},
 doi = {10.1086/516730},
 eprint = {astro-ph/0702626},
 journal = {\apj},
 month = {June},
 number = {1},
 pages = {627-641},
 primaryclass = {astro-ph},
 title = {{Protoplanetary Disk Turbulence Driven by the Streaming Instability: Nonlinear Saturation and Particle Concentration}},
 volume = {662},
 year = {2007}
}

@ARTICLE{Kirchschlager_2016,
       author = {{Kirchschlager}, Florian and {Wolf}, Sebastian and {Madlener}, David},
        title = "{The circumstellar disc of FS Tau B - a self-consistent model based on observations in the mid-infrared with NACO}",
      journal = {\mnras},
         year = 2016,
        month = oct,
       volume = {462},
       number = {1},
        pages = {858-866},
          doi = {10.1093/mnras/stw1692},
archivePrefix = {arXiv},
       eprint = {1607.05449},
 primaryClass = {astro-ph.SR},
       adsurl = {https://ui.adsabs.harvard.edu/abs/2016MNRAS.462..858K}
}

@ARTICLE{Laibe_2014,
       author = {{Laibe}, Guillaume and {Gonzalez}, Jean-Fran{\c{c}}ois and {Maddison}, Sarah T. and {Crespe}, Elisabeth},
        title = "{Growing dust grains in protoplanetary discs - III. Vertical settling}",
      journal = {\mnras},
         year = 2014,
        month = feb,
       volume = {437},
       number = {4},
        pages = {3055-3062},
          doi = {10.1093/mnras/stt1929},
archivePrefix = {arXiv},
       eprint = {1310.2638},
 primaryClass = {astro-ph.EP},
       adsurl = {https://ui.adsabs.harvard.edu/abs/2014MNRAS.437.3055L}
}

@article{Lambrechts2012,
 adsurl = {https://ui.adsabs.harvard.edu/abs/2012A&A...544A..32L},
 archiveprefix = {arXiv},
 author = {{Lambrechts}, M. and {Johansen}, A.},
 doi = {10.1051/0004-6361/201219127},
 eid = {A32},
 eprint = {1205.3030},
 journal = {\aap},
 month = {August},
 pages = {A32},
 primaryclass = {astro-ph.EP},
 title = {{Rapid growth of gas-giant cores by pebble accretion}},
 volume = {544},
 year = {2012}
}

@ARTICLE{Long2022,
       author = {{Long}, Feng and {Andrews}, Sean M. and {Rosotti}, Giovanni and {Harsono}, Daniel and {Pinilla}, Paola and {Wilner}, David J. and {{\"O}berg}, Karin I. and {Teague}, Richard and {Trapman}, Leon and {Tabone}, Beno{\^\i}t},
        title = "{Gas Disk Sizes from CO Line Observations: A Test of Angular Momentum Evolution}",
      journal = {\apj},
         year = 2022,
        month = may,
       volume = {931},
       number = {1},
          eid = {6},
        pages = {6},
          doi = {10.3847/1538-4357/ac634e},
archivePrefix = {arXiv},
       eprint = {2203.16735},
 primaryClass = {astro-ph.EP},
       adsurl = {https://ui.adsabs.harvard.edu/abs/2022ApJ...931....6L}
}

@ARTICLE{Louvet_2018,
       author = {{Louvet}, F. and {Dougados}, C. and {Cabrit}, S. and {Mardones}, D. and {M{\'e}nard}, F. and {Tabone}, B. and {Pinte}, C. and {Dent}, W.~R.~F.},
        title = "{The HH30 edge-on T Tauri star. A rotating and precessing monopolar outflow scrutinized by ALMA}",
      journal = {\aap},
         year = 2018,
        month = oct,
       volume = {618},
          eid = {A120},
        pages = {A120},
          doi = {10.1051/0004-6361/201731733},
archivePrefix = {arXiv},
       eprint = {1808.03285},
 primaryClass = {astro-ph.GA},
       adsurl = {https://ui.adsabs.harvard.edu/abs/2018A&A...618A.120L}
}

@ARTICLE{Luhman_2007,
       author = {{Luhman}, K.~L.},
        title = "{The Stellar Population of the Chamaeleon I Star-forming Region}",
      journal = {\apjs},
         year = 2007,
        month = nov,
       volume = {173},
       number = {1},
        pages = {104-136},
          doi = {10.1086/520114},
archivePrefix = {arXiv},
       eprint = {0710.3037},
 primaryClass = {astro-ph},
       adsurl = {https://ui.adsabs.harvard.edu/abs/2007ApJS..173..104L}
}

@ARTICLE{Luhman2009,
       author = {{Luhman}, K.~L. and {Mamajek}, E.~E. and {Allen}, P.~R. and {Cruz}, K.~L.},
        title = "{An Infrared/X-Ray Survey for New Members of the Taurus Star-Forming Region}",
      journal = {\apj},
         year = 2009,
        month = sep,
       volume = {703},
       number = {1},
        pages = {399-419},
          doi = {10.1088/0004-637X/703/1/399},
archivePrefix = {arXiv},
       eprint = {0911.5451},
 primaryClass = {astro-ph.GA},
       adsurl = {https://ui.adsabs.harvard.edu/abs/2009ApJ...703..399L}
}

@ARTICLE{Luhman_2010,
       author = {{Luhman}, K.~L. and {Allen}, P.~R. and {Espaillat}, C. and {Hartmann}, L. and {Calvet}, N.},
        title = "{The Disk Population of the Taurus Star-Forming Region}",
      journal = {\apjs},
         year = 2010,
        month = jan,
       volume = {186},
       number = {1},
        pages = {111-174},
          doi = {10.1088/0067-0049/186/1/111},
archivePrefix = {arXiv},
       eprint = {0911.5457},
 primaryClass = {astro-ph.GA},
       adsurl = {https://ui.adsabs.harvard.edu/abs/2010ApJS..186..111L}
}

@ARTICLE{Manara_2019,
       author = {{Manara}, C.~F. and {Tazzari}, M. and {Long}, F. and {Herczeg}, G.~J. and {Lodato}, G. and {Rota}, A.~A. and {Cazzoletti}, P. and {van der Plas}, G. and {Pinilla}, P. and {Dipierro}, G. and {Edwards}, S. and {Harsono}, D. and {Johnstone}, D. and {Liu}, Y. and {Menard}, F. and {Nisini}, B. and {Ragusa}, E. and {Boehler}, Y. and {Cabrit}, S.},
        title = "{Observational constraints on dust disk sizes in tidally truncated protoplanetary disks in multiple systems in the Taurus region}",
      journal = {\aap},
         year = 2019,
        month = aug,
       volume = {628},
          eid = {A95},
        pages = {A95},
          doi = {10.1051/0004-6361/201935964},
archivePrefix = {arXiv},
       eprint = {1907.03846},
 primaryClass = {astro-ph.EP},
       adsurl = {https://ui.adsabs.harvard.edu/abs/2019A&A...628A..95M}
}

@ARTICLE{Martinien_2024,
       author = {{Martinien}, L. and {M{\'e}nard}, F. and {Duch{\^e}ne}, G. and {Tazaki}, R. and {Perrin}, M.~D. and {Stapelfeldt}, K.~R. and {Pinte}, C. and {Wolff}, S.~G. and {Grady}, C. and {Dominik}, C. and {Roumesy}, M. and {Ma}, J. and {Ginski}, C. and {Benisty}, M. and {Hines}, D.~C. and {Schneider}, G.},
        title = "{The grazing-angle icy protoplanetary disk PDS 453}",
      journal = {\aap},
         year = 2024,
        month = dec,
       volume = {692},
          eid = {A111},
        pages = {A111},
          doi = {10.1051/0004-6361/202451475},
       adsurl = {https://ui.adsabs.harvard.edu/abs/2024A&A...692A.111M}
}

@ARTICLE{Martinien_2025a,
       author = {{Martinien}, L. and {Duch{\^e}ne}, G. and {M{\'e}nard}, F. and {Tazaki}, R. and {Stapelfeldt}, K.~R.},
        title = "{The role of absorption and scattering in shaping ice bands: Spatially resolved spectroscopy of protoplanetary disks}",
      journal = {\aap},
         year = 2025,
        month = apr,
       volume = {696},
          eid = {L6},
        pages = {L6},
          doi = {10.1051/0004-6361/202554360},
archivePrefix = {arXiv},
       eprint = {2503.16223},
 primaryClass = {astro-ph.EP},
       adsurl = {https://ui.adsabs.harvard.edu/abs/2025A&A...696L...6M}
}

@ARTICLE{Martinien_2025b,
       author = {{Martinien}, L. and {Duch{\^e}ne}, G. and {M{\'e}nard}, F. and {Stapelfeldt}, K.~R. and {Tazaki}, R. and {Bergner}, J.~B. and {Dartois}, E. and {Noble}, J.~A. and {Thompson}, W.~E.},
        title = "{Variation in the disk thickness across ice bands: A method for determining ice abundances in highly inclined protoplanetary disks}",
      journal = {\aap},
         year = 2025,
        month = dec,
       volume = {704},
          eid = {L17},
        pages = {L17},
          doi = {10.1051/0004-6361/202557328},
archivePrefix = {arXiv},
       eprint = {2510.11359},
 primaryClass = {astro-ph.SR},
       adsurl = {https://ui.adsabs.harvard.edu/abs/2025A&A...704L..17M}
}

@ARTICLE{McCabe_2011,
       author = {{McCabe}, C. and {Duch{\^e}ne}, G. and {Pinte}, C. and {Stapelfeldt}, K.~R. and {Ghez}, A.~M. and {M{\'e}nard}, F.},
        title = "{Spatially Resolving the HK Tau B Edge-on Disk from 1.2 to 4.7 {\ensuremath{\mu}}m: A Unique Scattered Light Disk}",
      journal = {\apj},
         year = 2011,
        month = feb,
       volume = {727},
       number = {2},
          eid = {90},
        pages = {90},
          doi = {10.1088/0004-637X/727/2/90},
       adsurl = {https://ui.adsabs.harvard.edu/abs/2011ApJ...727...90M}
}

@ARTICLE{Monin_1998,
       author = {{Monin}, J. -L. and {Menard}, F. and {Duchene}, G.},
        title = "{Using polarimetry to check rotation alignment in PMS binary stars. Principles of the method and first results}",
      journal = {\aap},
         year = 1998,
        month = nov,
       volume = {339},
        pages = {113-122},
       adsurl = {https://ui.adsabs.harvard.edu/abs/1998A&A...339..113M}
}

@ARTICLE{Muzic_2014,
       author = {{Mu{\v{z}}i{\'c}}, Koraljka and {Scholz}, Alexander and {Geers}, Vincent C. and {Jayawardhana}, Ray and {L{\'o}pez Mart{\'\i}}, Bel{\'e}n},
        title = "{Substellar Objects in Nearby Young Clusters (SONYC). VIII. Substellar Population in Lupus 3}",
      journal = {\apj},
         year = 2014,
        month = apr,
       volume = {785},
       number = {2},
          eid = {159},
        pages = {159},
          doi = {10.1088/0004-637X/785/2/159},
archivePrefix = {arXiv},
       eprint = {1403.0813},
 primaryClass = {astro-ph.GA},
       adsurl = {https://ui.adsabs.harvard.edu/abs/2014ApJ...785..159M}
}

@article{Perrin_2006,
 adsurl = {https://ui.adsabs.harvard.edu/abs/2006ApJ...645.1272P},
 archiveprefix = {arXiv},
 author = {{Perrin}, Marshall D. and {Duch{\^e}ne}, Gaspard and {Kalas}, Paul and {Graham}, James R.},
 doi = {10.1086/504510},
 eprint = {astro-ph/0603667},
 journal = {\apj},
 month = {July},
 number = {2},
 pages = {1272-1282},
 primaryclass = {astro-ph},
 title = {{Discovery of an Optically Thick, Edge-on Disk around the Herbig Ae Star PDS 144N}},
 volume = {645},
 year = {2006}
}

@ARTICLE{Pietu_2007,
       author = {{Pi{\'e}tu}, V. and {Dutrey}, A. and {Guilloteau}, S.},
        title = "{Probing the structure of protoplanetary disks: a comparative study of DM Tau, LkCa 15, and MWC 480}",
      journal = {\aap},
         year = 2007,
        month = may,
       volume = {467},
       number = {1},
        pages = {163-178},
          doi = {10.1051/0004-6361:20066537},
archivePrefix = {arXiv},
       eprint = {astro-ph/0701425},
 primaryClass = {astro-ph},
       adsurl = {https://ui.adsabs.harvard.edu/abs/2007A&A...467..163P}
}

@article{Pinte2018,
 adsurl = {https://ui.adsabs.harvard.edu/abs/2018ApJ...860L..13P},
 archiveprefix = {arXiv},
 author = {{Pinte}, C. and {Price}, D.~J. and {M{\'e}nard}, F. and {Duch{\^e}ne}, G. and {Dent}, W.~R.~F. and {Hill}, T. and {de Gregorio-Monsalvo}, I. and {Hales}, A. and {Mentiplay}, D.},
 doi = {10.3847/2041-8213/aac6dc},
 eid = {L13},
 eprint = {1805.10293},
 journal = {\apjl},
 month = {June},
 number = {1},
 pages = {L13},
 primaryclass = {astro-ph.SR},
 title = {{Kinematic Evidence for an Embedded Protoplanet in a Circumstellar Disk}},
 volume = {860},
 year = {2018}
}

@INPROCEEDINGS{Pinte2023,
       author = {{Pinte}, C. and {Teague}, R. and {Flaherty}, K. and {Hall}, C. and {Facchini}, S. and {Casassus}, S.},
        title = "{Kinematic Structures in Planet-Forming Disks}",
    booktitle = {Protostars and Planets VII},
         year = 2023,
       editor = {{Inutsuka}, S. and {Aikawa}, Y. and {Muto}, T. and {Tomida}, K. and {Tamura}, M.},
       series = {Astronomical Society of the Pacific Conference Series},
       volume = {534},
        month = jul,
        pages = {645},
          doi = {10.48550/arXiv.2203.09528},
archivePrefix = {arXiv},
       eprint = {2203.09528},
 primaryClass = {astro-ph.EP},
       adsurl = {https://ui.adsabs.harvard.edu/abs/2023ASPC..534..645P}
}

@article{Pollack1996,
 adsurl = {https://ui.adsabs.harvard.edu/abs/1996Icar..124...62P},
 author = {{Pollack}, James B. and {Hubickyj}, Olenka and {Bodenheimer}, Peter and {Lissauer}, Jack J. and {Podolak}, Morris and {Greenzweig}, Yuval},
 doi = {10.1006/icar.1996.0190},
 journal = {\icarus},
 month = {November},
 number = {1},
 pages = {62-85},
 title = {{Formation of the Giant Planets by Concurrent Accretion of Solids and Gas}},
 volume = {124},
 year = {1996}
}

@ARTICLE{Ribas2015,
       author = {{Ribas}, {\'A}lvaro and {Bouy}, Herv{\'e} and {Mer{\'\i}n}, Bruno},
        title = "{Protoplanetary disk lifetimes vs. stellar mass and possible implications for giant planet populations}",
      journal = {\aap},
         year = 2015,
        month = apr,
       volume = {576},
          eid = {A52},
        pages = {A52},
          doi = {10.1051/0004-6361/201424846},
archivePrefix = {arXiv},
       eprint = {1502.00631},
 primaryClass = {astro-ph.SR},
       adsurl = {https://ui.adsabs.harvard.edu/abs/2015A&A...576A..52R}
}

@ARTICLE{Rosenfeld2012,
       author = {{Rosenfeld}, Katherine A. and {Andrews}, Sean M. and {Wilner}, David J. and {Stempels}, H.~C.},
        title = "{A Disk-based Dynamical Mass Estimate for the Young Binary V4046 Sgr}",
      journal = {\apj},
         year = 2012,
        month = nov,
       volume = {759},
       number = {2},
          eid = {119},
        pages = {119},
          doi = {10.1088/0004-637X/759/2/119},
archivePrefix = {arXiv},
       eprint = {1209.4407},
 primaryClass = {astro-ph.SR},
       adsurl = {https://ui.adsabs.harvard.edu/abs/2012ApJ...759..119R}
}

@ARTICLE{Rota_2022,
       author = {{Rota}, A.~A. and {Manara}, C.~F. and {Miotello}, A. and {Lodato}, G. and {Facchini}, S. and {Koutoulaki}, M. and {Herczeg}, G. and {Long}, F. and {Tazzari}, M. and {Cabrit}, S. and {Harsono}, D. and {M{\'e}nard}, F. and {Pinilla}, P. and {van der Plas}, G. and {Ragusa}, E. and {Yen}, H.-W.},
        title = "{Observational constraints on gas disc sizes in the protoplanetary discs of multiple systems in the Taurus region}",
      journal = {\aap},
         year = 2022,
        month = jun,
       volume = {662},
          eid = {A121},
        pages = {A121},
          doi = {10.1051/0004-6361/202141035},
archivePrefix = {arXiv},
       eprint = {2201.03588},
 primaryClass = {astro-ph.EP},
       adsurl = {https://ui.adsabs.harvard.edu/abs/2022A&A...662A.121R}
}

@ARTICLE{Sanchis_2021,
       author = {{Sanchis}, E. and {Testi}, L. and {Natta}, A. and {Facchini}, S. and {Manara}, C.~F. and {Miotello}, A. and {Ercolano}, B. and {Henning}, Th. and {Preibisch}, T. and {Carpenter}, J.~M. and {de Gregorio-Monsalvo}, I. and {Jayawardhana}, R. and {Lopez}, C. and {Mu{\v{z}}i{\'c}}, K. and {Pascucci}, I. and {Santamar{\'\i}a-Miranda}, A. and {van Terwisga}, S. and {Williams}, J.~P.},
        title = "{Measuring the ratio of the gas and dust emission radii of protoplanetary disks in the Lupus star-forming region}",
      journal = {\aap},
         year = 2021,
        month = may,
       volume = {649},
          eid = {A19},
        pages = {A19},
          doi = {10.1051/0004-6361/202039733},
archivePrefix = {arXiv},
       eprint = {2101.11307},
 primaryClass = {astro-ph.EP},
       adsurl = {https://ui.adsabs.harvard.edu/abs/2021A&A...649A..19S}
}

@ARTICLE{Simon_2019,
       author = {{Simon}, M. and {Guilloteau}, S. and {Beck}, Tracy L. and {Chapillon}, E. and {Di Folco}, E. and {Dutrey}, A. and {Feiden}, Gregory A. and {Grosso}, N. and {Pi{\'e}tu}, V. and {Prato}, L. and {Schaefer}, Gail H.},
        title = "{Masses and Implications for Ages of Low-mass Pre-main-sequence Stars in Taurus and Ophiuchus}",
      journal = {\apj},
         year = 2019,
        month = oct,
       volume = {884},
       number = {1},
          eid = {42},
        pages = {42},
          doi = {10.3847/1538-4357/ab3e3b},
archivePrefix = {arXiv},
       eprint = {1908.10952},
 primaryClass = {astro-ph.SR},
       adsurl = {https://ui.adsabs.harvard.edu/abs/2019ApJ...884...42S}
}

@article{Stapelfeldt1998,
 adsurl = {http://adsabs.harvard.edu/abs/1998ApJ...502L..65S},
 author = {{Stapelfeldt}, K.~R. and {Krist}, J.~E. and {M{\'e}nard}, F. and {Bouvier}, J. and {Padgett}, D.~L. and {Burrows}, C.~J.},
 doi = {10.1086/311479},
 journal = {\apjl},
 month = {July},
 pages = {L65-L69},
 title = {{An Edge-On Circumstellar Disk in the Young Binary System HK Tauri}},
 volume = {502},
 year = {1998}
}

@INPROCEEDINGS{Stapelfeldt_2014,
       author = {{Stapelfeldt}, K.~R. and {Duch{\^e}ne}, G. and {Perrin}, M. and {Wolff}, S. and {Krist}, J.~E. and {Padgett}, D.~L. and {M{\'e}nard}, F. and {Pinte}, C.},
        title = "{HST Imaging of New Edge-on Circumstellar Disks in Nearby Star-forming Regions}",
    booktitle = {Exploring the Formation and Evolution of Planetary Systems},
         year = 2014,
       editor = {{Booth}, Mark and {Matthews}, Brenda C. and {Graham}, James R.},
       series = {IAU Symposium},
       volume = {299},
        month = jan,
        pages = {99-103},
          doi = {10.1017/S1743921313008004},
       adsurl = {https://ui.adsabs.harvard.edu/abs/2014IAUS..299...99S}
}

@article{Sturm2023_HH48_modeling,
 adsurl = {https://ui.adsabs.harvard.edu/abs/2023A&A...677A..17S},
 archiveprefix = {arXiv},
 author = {{Sturm}, J.~A. and {McClure}, M.~K. and {Law}, C.~J. and {Harsono}, D. and {Bergner}, J.~B. and {Dartois}, E. and {Drozdovskaya}, M.~N. and {Ioppolo}, S. and {{\"O}berg}, K.~I. and {Palumbo}, M.~E. and {Pendleton}, Y.~J. and {Rocha}, W.~R.~M. and {Terada}, H. and {Urso}, R.~G.},
 doi = {10.1051/0004-6361/202346052},
 eid = {A17},
 eprint = {2305.02338},
 journal = {\aap},
 month = {September},
 pages = {A17},
 primaryclass = {astro-ph.EP},
 title = {{The edge-on protoplanetary disk HH 48 NE. I. Modeling the geometry and stellar parameters}},
 volume = {677},
 year = {2023}
}

@ARTICLE{Tazaki_2025,
       author = {{Tazaki}, Ryo and {M{\'e}nard}, Fran{\c{c}}ois and {Duch{\^e}ne}, Gaspard and {Villenave}, Marion and {Ribas}, {\'A}lvaro and {Stapelfeldt}, Karl R. and {Perrin}, Marshall D. and {Pinte}, Christophe and {Wolff}, Schuyler G. and {Padgett}, Deborah L. and {Ma}, Jie and {Martinien}, Laurine and {Roumesy}, Maxime},
        title = "{JWST Imaging of Edge-on Protoplanetary Disks. IV. Mid-infrared Dust Scattering in the HH 30 Disk}",
      journal = {\apj},
         year = 2025,
        month = feb,
       volume = {980},
       number = {1},
          eid = {49},
        pages = {49},
          doi = {10.3847/1538-4357/ad9c6f},
archivePrefix = {arXiv},
       eprint = {2412.07523},
 primaryClass = {astro-ph.EP},
       adsurl = {https://ui.adsabs.harvard.edu/abs/2025ApJ...980...49T}
}

@ARTICLE{Teague_2018,
       author = {{Teague}, Richard and {Foreman-Mackey}, Daniel},
        title = "{A Robust Method to Measure Centroids of Spectral Lines}",
      journal = {Research Notes of the American Astronomical Society},
         year = 2018,
        month = Sep,
       volume = {2},
          eid = {173},
        pages = {173},
          doi = {10.3847/2515-5172/aae265},
       adsurl = {https://ui.adsabs.harvard.edu/abs/2018RNAAS...2c.173T}
}

@ARTICLE{Trapman_2019,
       author = {{Trapman}, L. and {Facchini}, S. and {Hogerheijde}, M.~R. and {van Dishoeck}, E.~F. and {Bruderer}, S.},
        title = "{Gas versus dust sizes of protoplanetary discs: effects of dust evolution}",
      journal = {\aap},
         year = 2019,
        month = sep,
       volume = {629},
          eid = {A79},
        pages = {A79},
          doi = {10.1051/0004-6361/201834723},
archivePrefix = {arXiv},
       eprint = {1903.06190},
 primaryClass = {astro-ph.EP},
       adsurl = {https://ui.adsabs.harvard.edu/abs/2019A&A...629A..79T}
}

@ARTICLE{Trapman_2025,
       author = {{Trapman}, Leon and {Vioque}, Miguel and {Kurtovic}, Nicol{\'a}s T. and {Zhang}, Ke and {Rosotti}, Giovanni P. and {Pinilla}, Paola and {Carpenter}, John and {Cieza}, Lucas A. and {Pascucci}, Ilaria and {Anania}, Rossella and {Agurto-Gangas}, Carolina and {Deng}, Dingshan and {Miley}, James and {P{\'e}rez}, Laura M. and {Sierra}, Anibal and {Tabone}, Beno{\^\i}t and {Ruiz-Rodriguez}, Dary A. and {Gonz{\'a}lez-Ruilova}, Camilo and {TorresVillanueva}, Estephani},
        title = "{The ALMA Survey of Gas Evolution of PROtoplanetary Disks (AGE-PRO). XI. Beam-corrected Gas Disk Sizes from Fitting $^{12}$CO Moment Zero Maps}",
      journal = {\apj},
         year = 2025,
        month = aug,
       volume = {989},
       number = {1},
          eid = {10},
        pages = {10},
          doi = {10.3847/1538-4357/adc7af},
archivePrefix = {arXiv},
       eprint = {2506.10750},
 primaryClass = {astro-ph.EP},
       adsurl = {https://ui.adsabs.harvard.edu/abs/2025ApJ...989...10T}
}

@ARTICLE{Tripathi2017,
       author = {{Tripathi}, Anjali and {Andrews}, Sean M. and {Birnstiel}, Tilman and {Wilner}, David J.},
        title = "{A millimeter Continuum Size-Luminosity Relationship for Protoplanetary Disks}",
      journal = {\apj},
         year = 2017,
        month = aug,
       volume = {845},
       number = {1},
          eid = {44},
        pages = {44},
          doi = {10.3847/1538-4357/aa7c62},
archivePrefix = {arXiv},
       eprint = {1706.08977},
 primaryClass = {astro-ph.EP},
       adsurl = {https://ui.adsabs.harvard.edu/abs/2017ApJ...845...44T}
}

@article{Vieira2003,
 adsurl = {https://ui.adsabs.harvard.edu/abs/2003AJ....126.2971V},
 author = {{Vieira}, S.~L.~A. and {Corradi}, W.~J.~B. and {Alencar}, S.~H.~P. and {Mendes}, L.~T.~S. and {Torres}, C.~A.~O. and {Quast}, G.~R. and {Guimar{\~a}es}, M.~M. and {da Silva}, L.},
 doi = {10.1086/379553},
 journal = {\aj},
 month = {December},
 number = {6},
 pages = {2971-2987},
 title = {{Investigation of 131 Herbig Ae/Be Candidate Stars}},
 volume = {126},
 year = {2003}
}

@ARTICLE{Villenave_2022,
       author = {{Villenave}, M. and {Stapelfeldt}, K.~R. and {Duch{\^e}ne}, G. and {M{\'e}nard}, F. and {Lambrechts}, M. and {Sierra}, A. and {Flores}, C. and {Dent}, W.~R.~F. and {Wolff}, S. and {Ribas}, {\'A}. and {Benisty}, M. and {Cuello}, N. and {Pinte}, C.},
        title = "{A Highly Settled Disk around Oph163131}",
      journal = {\apj},
         year = 2022,
        month = may,
       volume = {930},
       number = {1},
          eid = {11},
        pages = {11},
          doi = {10.3847/1538-4357/ac5fae},
archivePrefix = {arXiv},
       eprint = {2204.00640},
 primaryClass = {astro-ph.SR},
       adsurl = {https://ui.adsabs.harvard.edu/abs/2022ApJ...930...11V}
}

@ARTICLE{Villenave2020,
       author = {{Villenave}, M. and {M{\'e}nard}, F. and {Dent}, W.~R.~F. and {Duch{\^e}ne}, G. and {Stapelfeldt}, K.~R. and {Benisty}, M. and {Boehler}, Y. and {van der Plas}, G. and {Pinte}, C. and {Telkamp}, Z. and {Wolff}, S. and {Flores}, C. and {Lesur}, G. and {Louvet}, F. and {Riols}, A. and {Dougados}, C. and {Williams}, H. and {Padgett}, D.},
        title = "{Observations of edge-on protoplanetary disks with ALMA. I. Results from continuum data}",
      journal = {\aap},
         year = 2020,
        month = oct,
       volume = {642},
          eid = {A164},
        pages = {A164},
          doi = {10.1051/0004-6361/202038087},
archivePrefix = {arXiv},
       eprint = {2008.06518},
 primaryClass = {astro-ph.SR},
       adsurl = {https://ui.adsabs.harvard.edu/abs/2020A&A...642A.164V}
}

@ARTICLE{Vioque_2025,
       author = {{Vioque}, Miguel and {Kurtovic}, Nicol{\'a}s T. and {Trapman}, Leon and {Sierra}, Anibal and {P{\'e}rez}, Laura M. and {Zhang}, Ke and {Curone}, Pietro and {Rosotti}, Giovanni P. and {Carpenter}, John and {Tabone}, Beno{\^\i}t and {Pinilla}, Paola and {Deng}, Dingshan and {Pascucci}, Ilaria and {Miley}, James and {Agurto-Gangas}, Carolina and {Cieza}, Lucas A. and {Anania}, Rossella and {Ruiz-Rodriguez}, Dary A. and {Gonz{\'a}lez-Ruilova}, Camilo and {TorresVillanueva}, Estephani E. and {Kuznetsova}, Aleksandra},
        title = "{The ALMA Survey of Gas Evolution of PROtoplanetary Disks (AGE-PRO). X. Dust Substructures, Disk Geometries, and Dust-disk Radii}",
      journal = {\apj},
         year = 2025,
        month = aug,
       volume = {989},
       number = {1},
          eid = {9},
        pages = {9},
          doi = {10.3847/1538-4357/adc7b0},
archivePrefix = {arXiv},
       eprint = {2506.10746},
 primaryClass = {astro-ph.EP},
       adsurl = {https://ui.adsabs.harvard.edu/abs/2025ApJ...989....9V}
}

@ARTICLE{Walker2004,
       author = {{Walker}, Christina and {Wood}, Kenneth and {Lada}, C.~J. and {Robitaille}, Thomas and {Bjorkman}, J.~E. and {Whitney}, Barbara},
        title = "{The structure of brown dwarf circumstellar discs}",
      journal = {\mnras},
         year = 2004,
        month = jun,
       volume = {351},
       number = {2},
        pages = {607-616},
          doi = {10.1111/j.1365-2966.2004.07807.x},
archivePrefix = {arXiv},
       eprint = {astro-ph/0403276},
 primaryClass = {astro-ph},
       adsurl = {https://ui.adsabs.harvard.edu/abs/2004MNRAS.351..607W}
}

@ARTICLE{Weidenschilling_1977,
       author = {{Weidenschilling}, S.~J.},
        title = "{The Distribution of Mass in the Planetary System and Solar Nebula}",
      journal = {\apss},
         year = 1977,
        month = sep,
       volume = {51},
       number = {1},
        pages = {153-158},
          doi = {10.1007/BF00642464},
       adsurl = {https://ui.adsabs.harvard.edu/abs/1977Ap&SS..51..153W}
}

@article{Wolff2017,
 author = {Schuyler G. Wolff and Marshall D. Perrin and Karl Stapelfeldt and Gaspard Duchêne and Francois Ménard and Deborah Padgett and Christophe Pinte and Laurent Pueyo and William J. Fischer},
 doi = {10.3847/1538-4357/aa9981},
 journal = {The Astrophysical Journal},
 month = {dec},
 number = {1},
 pages = {56},
 publisher = {The American Astronomical Society},
 title = {Hubble Space Telescope Scattered-light Imaging and Modeling of the Edge-on Protoplanetary Disk ESO-Hα 569},
 url = {https://dx.doi.org/10.3847/1538-4357/aa9981},
 volume = {851},
 year = {2017}
}

@article{Wolff2021,
 adsurl = {https://ui.adsabs.harvard.edu/abs/2021AJ....161..238W},
 archiveprefix = {arXiv},
 author = {{Wolff}, Schuyler G. and {Duch{\^e}ne}, Gaspard and {Stapelfeldt}, Karl R. and {M{\'e}nard}, Francois and {Flores}, Christian and {Padgett}, Deborah and {Pinte}, Christophe and {Villenave}, Marion and {van der Plas}, Gerrit and {Perrin}, Marshall D.},
 doi = {10.3847/1538-3881/abeb1d},
 eid = {238},
 eprint = {2103.02665},
 journal = {\aj},
 month = {May},
 number = {5},
 pages = {238},
 primaryclass = {astro-ph.SR},
 title = {{The Anatomy of an Unusual Edge-on Protoplanetary Disk. I. Dust Settling in a Cold Disk}},
 volume = {161},
 year = {2021}
}

@INPROCEEDINGS{Yokogawa_2002,
       author = {{Yokogawa}, S. and {Kitamura}, Y. and {Momose}, M. and {Kawabe}, R.},
        title = "{Deep $^{13}$CO(J=1--0) Imaging of the Protostar Haro 6-5B: Discovery of Rotational Motion in the Protostellar Disk}",
    booktitle = {8th Asian-Pacific Regional Meeting, Volume II},
         year = 2002,
       editor = {{Ikeuchi}, Satoru and {Hearnshaw}, John and {Hanawa}, Tomoyuki},
        month = jan,
        pages = {239-240},
       adsurl = {https://ui.adsabs.harvard.edu/abs/2002aprm.conf..239Y}
}

@article{Youdin2005,
 adsurl = {https://ui.adsabs.harvard.edu/abs/2005ApJ...620..459Y},
 archiveprefix = {arXiv},
 author = {{Youdin}, Andrew N. and {Goodman}, Jeremy},
 doi = {10.1086/426895},
 eprint = {astro-ph/0409263},
 journal = {\apj},
 month = {February},
 number = {1},
 pages = {459-469},
 primaryclass = {astro-ph},
 title = {{Streaming Instabilities in Protoplanetary Disks}},
 volume = {620},
 year = {2005}
}

@ARTICLE{Zagaria_2021,
       author = {{Zagaria}, Francesco and {Rosotti}, Giovanni P. and {Lodato}, Giuseppe},
        title = "{On dust evolution in planet-forming discs in binary systems - I. Theoretical and numerical modelling: radial drift is faster in binary discs}",
      journal = {\mnras},
         year = 2021,
        month = jun,
       volume = {504},
       number = {2},
        pages = {2235-2252},
          doi = {10.1093/mnras/stab985},
archivePrefix = {arXiv},
       eprint = {2104.03022},
 primaryClass = {astro-ph.EP},
       adsurl = {https://ui.adsabs.harvard.edu/abs/2021MNRAS.504.2235Z}
}

@ARTICLE{Zagaria_2023,
       author = {{Zagaria}, Francesco and {Facchini}, Stefano and {Miotello}, Anna and {Manara}, Carlo F. and {Toci}, Claudia and {Clarke}, Cathie J.},
        title = "{Testing protoplanetary disc evolution with CO fluxes. A proof of concept in Lupus and Upper Sco}",
      journal = {\aap},
         year = 2023,
        month = apr,
       volume = {672},
          eid = {L15},
        pages = {L15},
          doi = {10.1051/0004-6361/202346164},
archivePrefix = {arXiv},
       eprint = {2304.01760},
 primaryClass = {astro-ph.EP},
       adsurl = {https://ui.adsabs.harvard.edu/abs/2023A&A...672L..15Z}
}

@ARTICLE{Zallio_2026,
       author = {{Zallio}, Luigi and {Rosotti}, Giovanni P. and {Vioque}, Miguel and {Miotello}, Anna and {Andrews}, Sean M. and {Manara}, Carlo F. and {Carpenter}, John M. and {Empey}, Aaron and {Kurtovic}, Nicol{\'a}s T. and {Law}, Charles J. and {Longarini}, Cristiano and {Paneque-Carre{\~n}o}, Teresa and {Teague}, Richard and {Villenave}, Marion and {Yen}, Hsi-Wei and {Zagaria}, Francesco},
        title = "{The $^{12}$CO gas structures of protoplanetary disks in the Upper Scorpius region}",
      journal = {\aap},
         year = 2026,
        month = jan,
       volume = {705},
          eid = {A49},
        pages = {A49},
          doi = {10.1051/0004-6361/202557366},
archivePrefix = {arXiv},
       eprint = {2511.16734},
 primaryClass = {astro-ph.SR},
       adsurl = {https://ui.adsabs.harvard.edu/abs/2026A&A...705A..49Z}
}

@ARTICLE{Zucker_2020,
       author = {{Zucker}, Catherine and {Speagle}, Joshua S. and {Schlafly}, Edward F. and {Green}, Gregory M. and {Finkbeiner}, Douglas P. and {Goodman}, Alyssa and {Alves}, Jo{\~a}o},
        title = "{A compendium of distances to molecular clouds in the Star Formation Handbook}",
      journal = {\aap},
         year = 2020,
        month = jan,
       volume = {633},
          eid = {A51},
        pages = {A51},
          doi = {10.1051/0004-6361/201936145},
archivePrefix = {arXiv},
       eprint = {2001.00591},
 primaryClass = {astro-ph.GA},
       adsurl = {https://ui.adsabs.harvard.edu/abs/2020A&A...633A..51Z}
}

\begin{appendix}

\section{Beam sizes}

We present the beam sizes obtained for the observations in Table \ref{table:beam_sizes}. 

\begin{table}[h]
\caption{Beam sizes of our observations.}
\centering
\begin{tabular}{ccccc}
\hline\hline
Sources & \multicolumn{2}{c}{Continuum} & \multicolumn{2}{c}{{$^{12}$CO}} \\

 &  FWHM (\arcsec)  & PA (°) & FWHM (\arcsec) & PA (°)\\
\hline
Flying Saucer   & 0.14 $\times$ 0.07 & -77.5 & 0.14 $\times$ 0.07 & -77.3\\
PDS 144 N   & 0.16 $\times$ 0.06 & -75.6 & 0.18 $\times$ 0.07 & -73.9\\
Oph 162305  & 0.11 $\times$ 0.05 & -78.0 & 0.14 $\times$ 0.07 & -76.0\\
Lup 160703  & 0.10 $\times$ 0.07 & -80.1 & 0.12 $\times$ 0.08 & -80.6\\
Lup 160708  & 0.10 $\times$ 0.07 & -82.7 & 0.12 $\times$ 0.08 & -80.9\\
HK Tau B    & 0.11 $\times$ 0.07 & 3 & 0.13 $\times$ 0.12 & 79.7\\
HV Tau C    &  0.11 $\times$ 0.08 & 4 & 0.14 $\times$ 0.13 & 78.0\\
Haro 6-5B   &  0.12 $\times$ 0.07 & -19 & 0.15 $\times$ 0.14 & -32.8\\
Tau 042307  & 0.11 $\times$ 0.07 & -8 & 0.24 $\times$ 0.22 & -76.1\\
Tau 042021  & 0.12 $\times$ 0.08 & -13 & 0.26 $\times$ 0.24 & -77.8\\
ESO H$\alpha$ 569&  0.48 $\times$ 0.28 & -8 & 0.50 $\times$ 0.30 & -10.4\\
ESO H$\alpha$ 574&  0.47 $\times$ 0.28 & -10 & 0.49 $\times$ 0.30 & -11.7\\
HH 48 NE    &  0.48 $\times$ 0.28 & -5 & 0.52 $\times$ 0.31 & -8.3\\
Oph 163131  &  0.48 $\times$ 0.28 & -70 & 0.08 $\times$ 0.07 & 88.9\\
\hline
\end{tabular}
\tablefoot{All observations are in Band 7 except Oph 163131 which are in Band 6.}
\label{table:beam_sizes}
\end{table}

\section{Parameters of disk-hosting companions}
\label{sect:BINARIES}

We performed the same analyses on the disk-hosting companions of the multiple systems as we did for the objects in our main sample. The continuum and line fluxes are listed in Table \ref{table:FLUX_COMPANIONS}. The position angles and inclinations are summarized in Table. \ref{table:DISK_PARAMETERS_BINARIES}. The major axis sizes as well as the minor axis sizes are listed in Table \ref{table:SIZES_MAJOR_BINARIES} and \ref{table:SIZES_MINOR_BINARIES}, respectively. The PV diagrams are presented in Fig. \ref{fig:PV_DIAGRAMS_BINARIES}. Finally, the estimated systemic velocities as well as dynamical stellar masses are presented in Table \ref{table:MASSES_BINARIES}.

\begin{table}
\caption{ALMA fluxes of the companions.}
\centering
\begin{tabular}{ccc}
\hline\hline
Sources & Continuum flux & $^{12}$CO (3-2) line flux  \\
& (mJy) & (Jy km s$^{-1}$)\\
\hline
PDS 144 S   & 57.0 $\pm$ 5.7 & 2.1 $\pm$ 0.2\\
HK Tau A    & 74.6 $\pm$ 7.5 & 2.0 $\pm$ 0.2 \\
HH 48 SW    & 58.0 $\pm$ 5.8 & 4.6 $\pm$ 0.4\\
\hline
\end{tabular}
\label{table:FLUX_COMPANIONS}
\end{table}

\begin{table}[h!]
\caption{Disks parameters of the companions.}
\centering
\begin{threeparttable}
\begin{tabular}{cccc}
\hline\hline
Sources & PA & $i_{\mathrm{AxisRatio}}$ & $i_{literature}$ \\
& (\degr )& (\degr) & (\degr) \\
\hline  
PDS 144 S     & 35  & >74 & 73 $\pm$ 7$^{a}$ \\
HK Tau A      & 178 & >54 & 63.1 $\pm$ 1.5$^{b}$ \\
HH 48 SW      & 4   & >75 & - \\
\hline
\end{tabular}
\label{table:BINARIES_PARAMETERS}
\footnotesize
\begin{minipage}{\linewidth} 
\textbf{References}.
$^{a}$\citet{Hornbeck_2012}, 
$^{b}$\citet{Simon_2019}.
\end{minipage}
\end{threeparttable}
\label{table:DISK_PARAMETERS_BINARIES}
\end{table}

\begin{table}
\caption{Major axis diameters measurements from the ALMA continuum, ALMA $^{12}$CO and HST scattered light observations.}
\centering
\begin{tabular}{cccccccc}
\hline\hline
Sources & ALMA & HST &  ALMA  \\
& continuum & scattered & $^{12}$CO\\
  & (\arcsec) & (\arcsec) & (\arcsec) \\
\hline
PDS 144 S    & 0.85 $\pm$ 0.05 & unresolved & 1.38 $\pm$ 0.10\\
HK Tau A     & 0.61 $\pm$ 0.05 & unresolved & 1.33 $\pm$ 0.10  \\
HH 48 SW     & 1.37 $\pm$ 0.07 &  unresolved & 3.61 $\pm$ 0.11 \\
\hline
\end{tabular}
\label{table:SIZES_MAJOR_BINARIES}
\end{table}

\begin{table}
\caption{Minor axis sizes measurements from the ALMA continuum,  ALMA $^{12}$CO and HST-VLT scattered light observations.}
\centering
\begin{tabular}{cccccccc}
\hline\hline
Sources & ALMA & HST &  ALMA  \\
& continuum & scattered & $^{12}$CO\\
  & (\arcsec) & (\arcsec) & (\arcsec) \\
\hline
PDS 144 S    &  0.12 $\pm$ 0.03 & unresolved & 0.68 $\pm$ 0.02\\
HK Tau A     &  0.14 $\pm$ 0.01 & unresolved & 0.56  $\pm$ 0.02\\
HH 48 SW     &  0.17 $\pm$ 0.01 & unresolved & 1.64 $\pm$ 0.06 \\
\hline
\end{tabular}
\tablefoot{See Sect. \ref{sect:results_alma_vertical} and \ref{sect:results_hst_vertical} for definitions of the three quantities.}
\label{table:SIZES_MINOR_BINARIES}
\end{table}

\begin{figure}[h!]
        \centering
        \includegraphics[width=0.5\textwidth]{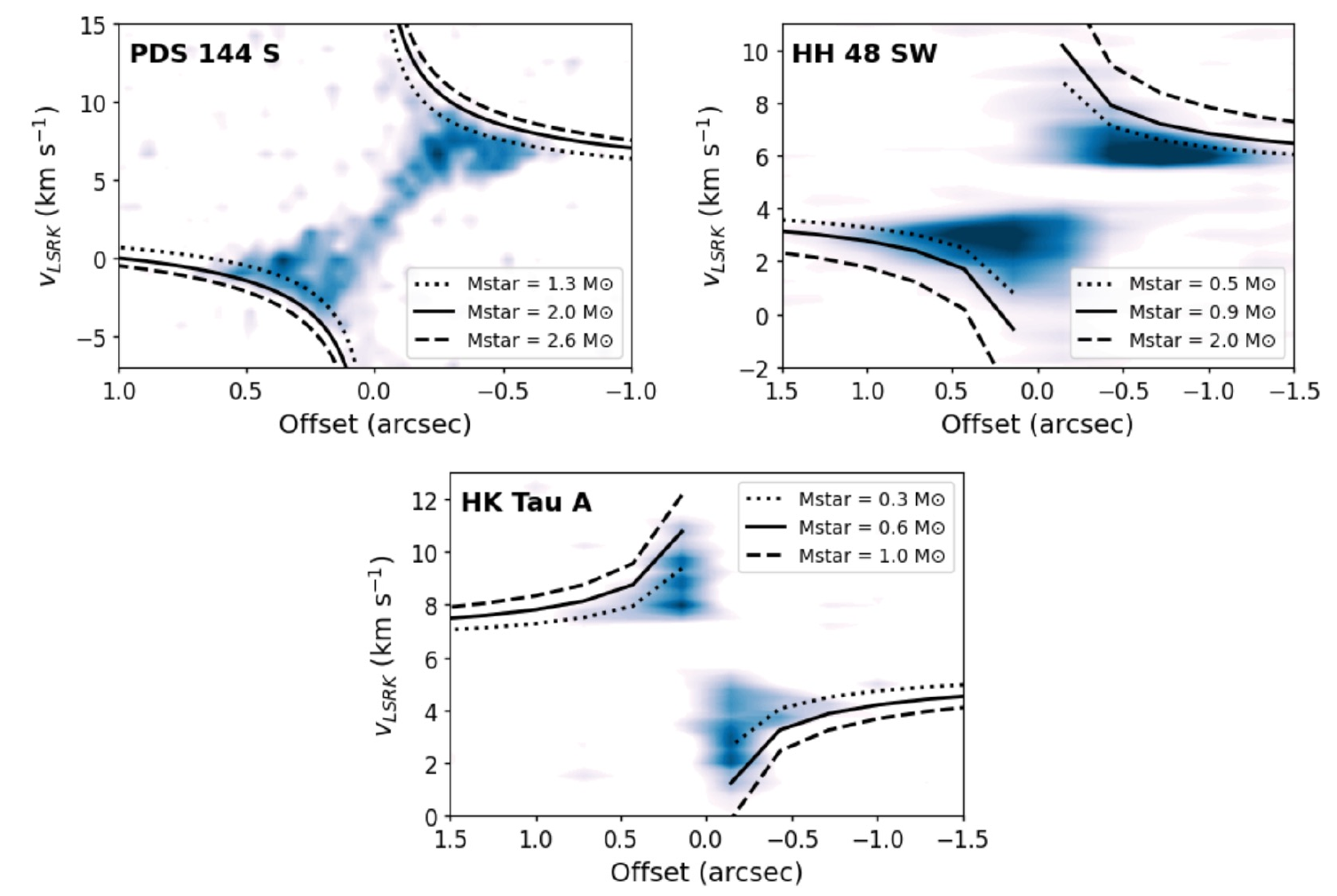}
        \caption{Position-velocity diagrams of the companions with 3 Keplerian velocity curves corresponding to the preferred dynamical mass of the star, as well as conservative upper and lower limits.}
        \label{fig:PV_DIAGRAMS_BINARIES}
\end{figure}

\begin{table}
\caption{Systemic velocities and dynamical masses from this work and from the literature.}
\centering
\begin{threeparttable}
\begin{tabular}{cccc}
\hline\hline
Sources & V$_{sys}$ & M$_{\star}$ & M$_{\star}$$_{literature}$ \\
  & (km s$^{-1}$) &  (M$\odot$) & (M$\odot$)\\
\hline
PDS 144 S & 3.5 & 2.0 (1.3..2.6) & - \\
HK Tau A  & 6.0 & 0.6 (0.3..1.0) & 0.53 $\pm$ 0.03$^{a}$\\
HH 48 SW  & 4.8 & 0.9 (0.5..2.0) & - \\
\hline
\end{tabular}
\begin{tablenotes}[flushleft]
\footnotesize
\item[]
\begin{minipage}{\linewidth} 
\textbf{References}.
$^{a}$\citet{Simon_2019}.
\end{minipage}
\end{tablenotes}
\end{threeparttable}
\label{table:MASSES_BINARIES}
\end{table}

\end{appendix}

\label{LastPage}
\end{document}